\documentclass[pre,twocolumn,aps,10pt]{revtex4-2}
\usepackage{amsmath}
\usepackage{amssymb}
\usepackage{graphicx}
\usepackage{braket}
\usepackage{stmaryrd}
\usepackage{hyperref}
\usepackage{color}
\usepackage{dsfont}
\usepackage{booktabs}
\usepackage{bm}

\hypersetup{
    colorlinks,
    citecolor=blue,
    linkcolor=blue,
    urlcolor=blue
}

\newcommand{\EndMatter}{
\vskip 0.5cm
\centerline{\textbf{END MATTER}}
\vskip 0.5cm
}

\newcommand{\concentrationUDAUratio}{(S30)}

\begin{document}

\title{Quantum Thermodynamic Uncertainty Relations without Quantum Corrections:\\ A Coherent-Incoherent Correspondence Approach}

\author{Tomohiro Nishiyama}
\email{htam0ybboh@gmail.com}
\affiliation{Independent Researcher, Tokyo 206-0003, Japan}

\author{Yoshihiko Hasegawa}
\email{hasegawa@biom.t.u-tokyo.ac.jp}
\affiliation{Department of Information and Communication Engineering, Graduate
School of Information Science and Technology, The University of Tokyo,
Tokyo 113-8656, Japan}

\date{\today}
\begin{abstract}

We introduce the coherent-incoherent correspondence as a framework for deriving quantum thermodynamic uncertainty relations under continuous measurement in Lindblad dynamics. 
The coherent-incoherent correspondence establishes a mapping between the original quantum system that undergoes \textit{ coherent} evolution and its corresponding \textit{incoherent} system without coherent dynamics.
The coherent-incoherent correspondence relates quantities across these two systems, including jump statistics, dynamical activity, and entropy production.
Since the classical-like properties of the incoherent system allow us to derive thermodynamic uncertainty relations within it,
these relations can be transferred to the coherent system via the coherent-incoherent correspondence.
This enables us to derive quantum thermodynamic uncertainty relations for the original coherent system.
Unlike existing quantum uncertainty relations, which typically require explicit quantum correction terms, our approach avoids these additional terms. 
This means that we can establish a lower bound for quantum entropy production using only current statistics. This approach opens up new possibilities for inferring entropy production in quantum systems.
Through numerical calculations for a model with coherent jump operators, we show that steady-state coherence lowers the bounds on precision (i.e., allows higher precision).

\end{abstract}
\maketitle

\begin{figure}
    \centering
    \includegraphics[width=1\linewidth]{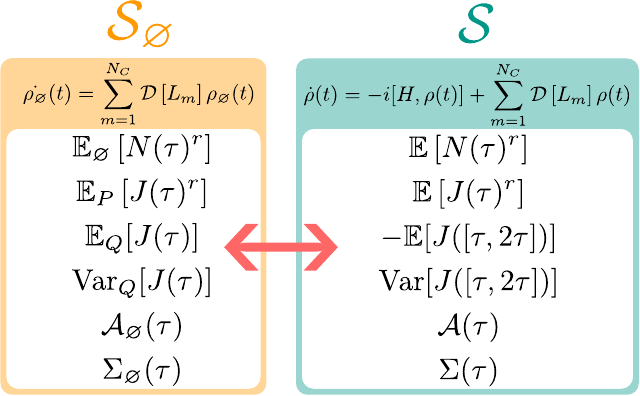}
    \caption{
    Illustration of the CIC. 
    There are correspondences of quantities between the coherent system $\mathcal{S}$ and incoherent system $\mathcal{S}_\varnothing$. 
These correspondences are applied to uncertainty relations in $\mathcal{S}_\varnothing$ to derive analogous relations in $\mathcal{S}$.
In this figure,
$N(\tau)$ represents a trajectory observable and $J(\tau)$ is a current observable. 
$\mathcal{A}_\varnothing$ and $\mathcal{A}$ represent the dynamical activity in $\mathcal{S}_\varnothing$ and $\mathcal{S}$, respectively; the same applies to the entropy production $\Sigma$.
$\mathbb{E}_\varnothing[\bullet]$ and $\mathbb{E}[\bullet]$ represent the expectation values in $\mathcal{S}_\varnothing$ and $\mathcal{S}$ respectively.
$\mathbb{E}_P[\bullet]$ and $\mathbb{E}_Q[\bullet]$ each represent the expectation values for forward and backward probabilities 
in $\mathcal{S}_\varnothing$. The same meaning applies to the variance $\mathrm{Var}[\bullet]$.
    }
    \label{fig:comparison}
\end{figure}

\textit{Introduction.---}
Achieving greater precision in thermal machines requires increased thermodynamic resources.  This fundamental fact is formulated as the thermodynamic uncertainty relation (TUR) \cite{Barato:2015:UncRel,Gingrich:2016:TUP} (see Ref.~\cite{Horowitz:2019:TURReview} for a review).  The TUR establishes a fundamental trade-off between the precision of an observable current and the associated thermodynamic cost, typically quantified by entropy production or dynamical activity. For classical systems, it often states that the relative variance of a current (variance divided by the squared mean) is bounded from below by twice the inverse of the entropy production or the inverse of the dynamical activity. 
Recently, TURs have attracted significant attention in quantum systems \cite{Erker:2017:QClockTUR,Brandner:2018:Transport,Carollo:2019:QuantumLDP,Liu:2019:QTUR,Guarnieri:2019:QTURPRR,Saryal:2019:TUR,Hasegawa:2020:QTURPRL,Hasegawa:2020:TUROQS,Kalaee:2021:QTURPRE,Monnai:2022:QTUR,Hasegawa:2023:BulkBoundaryBoundNC,Nishiyama:2024:NonHermiteQSLPRA,nishiyama2025speed, Prech:2025:CoherenceQTUR,vu2025fundamental,Moreira:2025:MultiTUR}. 
In particular, many studies have explored continuous measurement in Lindblad dynamics (see Ref.~\cite{Landi:2023:CurFlucReviewPRXQ} for a review), and it has been reported that precision can be enhanced due to quantum effects \cite{Hasegawa:2020:QTURPRL}.
In this Letter, we introduce a concept called the \textit{coherent-incoherent correspondence} (CIC) to derive quantum TURs under continuous measurement. The CIC establishes a correspondence between two systems: system $\mathcal{S}$, which is the original Lindblad dynamics and includes coherent dynamics, and system $\mathcal{S}_\varnothing$, which lacks coherent dynamics (Fig.~\ref{fig:comparison}). 
The CIC allows us to establish correspondences between quantities, such as jump statistics, dynamical activity, and entropy production, in $\mathcal{S}$ and $\mathcal{S}_\varnothing$.
Since $\mathcal{S}_\varnothing$ lacks coherent dynamics, its behavior resembles that of a classical Markov process, making it straightforward to derive trade-off relations. The CIC allows trade-off relations to be transferred from $\mathcal{S}_\varnothing$ to $\mathcal{S}$, which facilitates the derivation of quantum TURs that hold in the original Lindblad dynamics $\mathcal{S}$.
Specifically, we derive quantum TURs for the Lindblad dynamics, where the thermodynamic costs are given by the dynamical activity, a measure of the total number of stochastic jump events in the system's evolution  [cf. Eqs.~\eqref{eq:main_TUR} and \eqref{eq:main_TUR_diff}], and the entropy production [cf. Eq.~\eqref{eq:TUR_EP_main}] (see Table~\ref{tab:results} for the summary of results).

A significant aspect of the derived relations in $\mathcal{S}$ is that they do not include ``quantum correction terms''.
In quantum TURs based on dynamical activity, such as in Refs.~\cite{Hasegawa:2020:QTURPRL,Nishiyama:2024:ExactQDAPRE}, quantum coherent components are incorporated into the dynamical activity.
A recent study \cite{Prech:2025:CoherenceQTUR} derived a TUR for the dynamical activity, where the expectation value has an additional term arising from the coherent dynamics in place of an additional term in the dynamical activity. 
Regarding the entropy production, similar precision improvements due to coherent components have been pointed out in Ref.~\cite{vu2025fundamental},
where a correction term due to coherent dynamics is added to the expectation value.
Therefore, in previous studies, the thermodynamic cost part or the expectation value part included quantum correction terms derived from coherent dynamics.
While these corrections are crucial for the validity of the bounds, they often lack a clear physical interpretation and can be challenging to measure experimentally. 
This Letter provides a general framework for deriving trade-off relations in quantum systems by establishing correspondences between coherent and incoherent systems.

\textit{Methods.---}
We consider TURs in
an open quantum system whose dynamics is described by the Lindblad equation. The Lindblad equation characterizes how a quantum system evolves when it interacts with an external environment. Let $H$ represent the Hamiltonian of the system, and let $L_m$ denote the $m$-th jump operator, where $m$ ranges from 1 to $N_C$ (the total number of channels).
The Lindblad equation is given by
\begin{align}
    \dot{\rho}(t)=\mathcal{L}\rho(t)=-i\left[H,\rho(t)\right]+\sum_{m=1}^{N_{C}}\mathcal{D}\left[L_{m}\right]\rho(t),
    \label{eq:Lindblad_def}
\end{align}
where $\mathcal{L}$ is the Lindblad superoperator and 
$\mathcal{D}[L]\rho:= L\rho L^{\dagger}-\frac{1}{2}\{L^{\dagger}L,\rho\}$ 
is known as the dissipator, which describes the interaction between the system and its environment.

Throughout this study, we assume the following condition: for each $m$, there exists $\omega_m \in \mathbb{R}$ such that $H$ and $L_m$ satisfy
\begin{align}
    [L_m, H]=\omega_m L_m.
    \label{eq:condition}
\end{align}
Here, $\omega_m$ represents the energy difference between the initial and final states (i.e. the energy before minus the energy after the transition).
The condition~\eqref{eq:condition} and $[L_m^\dagger, H]=-\omega_m L_m^\dagger$ yield $[H, L_m^\dagger L_m]=0$.
The Lindblad equation can be derived by assuming the semigroup property of the dynamical map, which does not require Eq.\eqref{eq:condition}. However, when deriving the Lindblad equation from a microscopic model using the Born-Markov and secular approximations, Eq.\eqref{eq:condition} becomes necessary \cite{Breuer:2002:OpenQuantum}.
Under Eq.~\eqref{eq:condition},
the steady-state density operator $\rho_\mathrm{ss}$ satisfies
$[\rho_\mathrm{ss},H]=0$.

In the Lindblad equation, continuous measurement corresponds to continuously monitoring the environment coupled to the system. The result of this continuous monitoring is a measurement record that documents each jump event and the time at which it occurred.  
Suppose that we observe $K$ jump events in the time interval $[0,\tau]$, and let $m_k \in \{1,\cdots,N_C\}$ denote the type of the $k$-th jump at time $t_k$. In addition to continuous monitoring, we can also perform a measurement on the system itself at time $t=\tau$, yielding an outcome $s$. The complete measurement record can then be represented as
\begin{align}
    \zeta_\tau := \left[\left(t_1, m_1\right), \left(t_2, m_2\right), \ldots, \left(t_K, m_K\right);s\right].
    \label{eq:traj_zeta_def}
\end{align}
Here, the sequence $\zeta_\tau$ is referred to as a \textit{trajectory}.

We next define the observables for the continuous measurement. 
Let $N(\zeta_\tau)$ be an arbitrary observable of the trajectory $\zeta_\tau$  that depends only on $\{m_j\}$ and not on $\{t_j\}$.  
The observable $N(\zeta_\tau)$ is the key quantity in this Letter. Throughout this Letter, we simplify the notation $N(\zeta_\tau)$ to $N(\tau)$ when we do not need to write the trajectory explicitly. 
An instance of $N(\tau)$ considered in this study is the counting observable $N^{C}(\tau)$ defined by
\begin{align}
    N^{C}(\tau):=\sum_{m}c_{m}N_{m}(\tau),
    \label{eq:M_def}
\end{align}
where $[c_{m}]$ is a real weight vector and $N_m(\tau)$ is the number of the $m$-th jumps within $[0,\tau]$. Later, we will consider the thermodynamic current [cf. $J(\tau)$ in  Eq.~\eqref{eq:J_current_def}] for another instance of the trajectory observable.

\textit{Results.---}
This Letter proposes a concept referred to as the CIC to derive quantum TURs. 
The CIC establishes a relationship between two quantum systems. The first system, $\mathcal{S}$, follows the original Lindblad dynamics [Eq.~\eqref{eq:Lindblad_def}], which includes a coherent evolution through the term $-i[H,\rho]$. 
The second system, $\mathcal{S}_{\varnothing}$, is identical to $\mathcal{S}$ except that it lacks coherent evolution; it has the same jump operators and initial density operator as $\mathcal{S}$, but its Hamiltonian $H$ is set to zero.
Specifically, the system $\mathcal{S}_{\varnothing}$ is governed by 
\begin{align}
    \rho_{\varnothing}(0)&=\rho(0),  \label{eq:initial_def_S0}\\
    \dot{\rho}_{\varnothing}(t)&=\sum_{m=1}^{N_{C}}\mathcal{D}\left[L_{m}\right]\rho_{\varnothing}(t).
    \label{eq:Lindblad_def_S0}
\end{align}
The exclusion of the coherent contribution in $\mathcal{S}_\varnothing$ leads to a more tractable calculation of several quantities compared to the calculations performed using $\mathcal{S}$.
By comparing these two systems and transferring statistical information from $\mathcal{S}_{\varnothing}$ to $\mathcal{S}$, we can derive trade-off relations for $\mathcal{S}$ (Fig.~\ref{fig:comparison}).
In what follows, variables with the subscript $\varnothing$ denote quantities in the system $\mathcal{S}_{\varnothing}$.

 The physical intuition behind the CIC lies in the recognition that the incoherent system $\mathcal{S}_{\varnothing}$ is mathematically equivalent to the interaction picture of the original coherent system $\mathcal{S}$. 
Letting $U(t):=e^{-iHt}$,
it follows from 
Eq.~\eqref{eq:condition} that (see the End Matter)
\begin{align}
    L_m U(t)=e^{-i\omega_m t}U(t)L_m. \label{eq:UL_commute1}
\end{align}
Equation~\eqref{eq:UL_commute1} shows that
the coherent evolution driven by the Hamiltonian $H$ only applies a phase factor to the quantum state during the no-jump intervals. Since the probabilities of quantum jump trajectories are calculated from the squared magnitude of the state amplitude, these phase factors cancel out. This cancellation shows that the statistics of jump events are identical in both systems:
\begin{align}
    \mathbb{E}[N(\tau)^r]&=\mathbb{E}_{\varnothing}[N(\tau)^r],
    \label{eq:main_result_moment} \\
    \mathbb{E}[|N(\tau)|^r]&=\mathbb{E}_{\varnothing}[|N(\tau)|^r],\label{eq:main_result_moment_abs}
\end{align}    
where $r$ is a real number.
Equations~\eqref{eq:main_result_moment} and~\eqref{eq:main_result_moment_abs} are 
the key results in the CIC. 
The proofs of Eqs.~\eqref{eq:main_result_moment} and~\eqref{eq:main_result_moment_abs} are shown in the End Matter. 
Not all statistical quantities match between $\mathcal{S}$ and $\mathcal{S}_\varnothing$. For example, the fidelity $\mathrm{Fid}(\bullet,\bullet)$ between states depends on coherent dynamics, so in general
$\mathrm{Fid}(\rho(0),\rho(\tau)) \ne \mathrm{Fid}(\rho_\varnothing(0),\rho_\varnothing(\tau))$.

\textit{Dynamical activity case.}---
The dynamical activity is a thermodynamic quantity that characterizes the activity of the system and plays a central role in trade-off relations \cite{Garrahan:2017:TUR,Shiraishi:2018:SpeedLimit,Terlizzi:2019:KUR}. 
By substituting $N^{C}(\tau)=\sum_m N_m(\tau)$ and $r=1$ in Eq.~\eqref{eq:main_result_moment}, we obtain the CIC for the dynamical activity: 
\begin{align}
    \mathcal{A}(\tau)&:=\mathbb{E}\left[\sum_{m}N_{m}(\tau)\right]=\mathcal{A}_{\varnothing}(\tau),\label{eq:DA_S0}\\
    \mathfrak{a}(\tau)&:=\dot{\mathcal{A}}(\tau)=\mathfrak{a}_{\varnothing}(\tau).\label{eq:DA_ratio_S0}
\end{align}
Equations~\eqref{eq:main_result_moment}--\eqref{eq:DA_ratio_S0} imply that any trade-off relations in $\mathcal{S}_{\varnothing}$ involving the quantities $\mathbb{E}_{\varnothing}[N(\tau)^r]$, $\mathcal{A}_{\varnothing}(\tau)$ should also hold in the original system $\mathcal{S}$ for $\mathbb{E}[N(\tau)^r]$ and $\mathcal{A}(\tau)$. 
There is an uncertainty relation
involving $\mathrm{Var}[N(\tau)]$, $\mathbb{E}[N(\tau)]$, and $\mathcal{A}(\tau)$
that holds in $\mathcal{S}_\varnothing$ \cite{Hasegawa:2023:BulkBoundaryBoundNC}. 
By replacing the quantities in $\mathcal{S}_\varnothing$ with those in $\mathcal{S}$ with the CIC, we obtain
\begin{align}
\left(\frac{\sqrt{\mathrm{Var}[N(\tau_{2})]}+\sqrt{\mathrm{Var}[N(\tau_{1})]}}{\mathbb{E}[N(\tau_{2})]-\mathbb{E}[N(\tau_{1})]}\right)^{2}\geq\tan\left[\frac{1}{2}\int_{\tau_{1}}^{\tau_{2}}\frac{\sqrt{\mathcal{A}(t)}}{t}dt\right]^{-2},
    \label{eq:main_TUR}
\end{align}
where we assume $(1/2)\int_{\tau_1}^{\tau_2}\sqrt{\mathcal{A}(t)}/t\,dt\le\pi/2$.
When $\tau_2 = \tau$ and $\tau_1 = \tau - \epsilon$, where $\epsilon$ is infinitesimally small, Eq.~\eqref{eq:main_TUR} reduces to
\begin{align}
    \frac{\mathrm{Var}[N(\tau)]}{\tau^{2}\left(\partial_{\tau}\mathbb{E}[N(\tau)]\right)^{2}}\geq\frac{1}{\mathcal{A}(\tau)}.
    \label{eq:main_TUR_diff}
\end{align}
Equations~\eqref{eq:main_TUR} and \eqref{eq:main_TUR_diff} constitute the first result of this Letter. 
In Refs.~\cite{Hasegawa:2020:QTURPRL,Hasegawa:2023:BulkBoundaryBoundNC}, trade-off relations similar to Eqs.~\eqref{eq:main_TUR} and \eqref{eq:main_TUR_diff} were derived, where $\mathcal{A}(\tau)$ is replaced by the quantum dynamical activity $\mathcal{B}(\tau)\equiv \mathcal{A}(\tau) + \mathcal{C}(\tau)$,
where $\mathcal{C}(\tau)$ is a coherent term representing the effect of coherent dynamics of the Lindblad equation  (see the supplementary material \cite{Supp:2025:CIC} for the explicit expression).   
Note that Refs.~\cite{Hasegawa:2020:QTURPRL,Hasegawa:2023:BulkBoundaryBoundNC} do not require the assumption of Eq.~\eqref{eq:condition}.
Equations~\eqref{eq:main_TUR} and \eqref{eq:main_TUR_diff} mirror, in form, the classical TURs presented in \cite{Hasegawa:2023:BulkBoundaryBoundNC} and \cite{Terlizzi:2019:KUR}. Consequently, with the assumption given by Eq.~\eqref{eq:condition}, our results demonstrate that the analogous relations that hold in classical systems also hold in quantum systems. 

However, this equivalence does not imply that quantum effects can not be taken into account in the derived relations~\eqref{eq:main_TUR} and \eqref{eq:main_TUR_diff}.
This is because the initial density operator may contain coherence, non-diagonal elements with respect to the Hamiltonian, and the existence of such coherence alters the precision limit. 
When we consider the steady-state condition and $N(\tau)=N^C(\tau)$ in Eq.~\eqref{eq:main_TUR_diff}, 
\begin{align}
    \frac{\mathrm{Var}[N^{C}(\tau)]}{\mathbb{E}[N^{C}(\tau)]^{2}}\ge\frac{1}{\mathcal{A}(\tau)}=\frac{1}{\mathfrak{a}\tau},
    \label{eq:KUR_steadystate}
\end{align}
holds, which is identical to the bound in Ref.~\cite{Garrahan:2017:TUR}.

When the jump operators correspond to transitions between the eigenstates of $H$,
the same form as the classical TUR is known to hold in the transient regime even when the initial density operator contains coherence \cite{Vu:2021:QTURPRL}. 
Moreover, in such a case, 
the non-diagonal elements of the density operator vanish in the steady-state regime, which indicates that the system behaves essentially as a classical system. However, when the jump operators include coherent transitions \cite{Scully:2011:CoherentHE, Holubec:2018:QHE}, the steady-state density operator retains nonzero non-diagonal elements.
In this scenario, it has been unclear whether the bounds remain valid when the quantum corrections are excluded.
We argue that Eqs.~\eqref{eq:main_TUR}--\eqref{eq:KUR_steadystate} hold for the systems including coherent transitions as long as the condition of Eq.~\eqref{eq:condition} is satisfied, which is elaborated later.

\textit{Entropy production case.}---
Next, we consider entropy production, which plays a central role in uncertainty relations. 
We assume the local detailed balance condition \cite{Horowitz:2013:QJ}: 
\begin{align}
    L_{m}&=e^{\frac{\Delta s_m}{2}}L_{m^\prime}^\dagger,
    \label{eq:ldb}
\end{align}
where $\Delta s_m$ denotes the entropy change of the environment due to the jump such that $\Delta s_{m^\prime}=-\Delta s_m$. The entropy production is defined as 
\begin{align}
    \Sigma(\tau)&:=\mathrm{Tr}[\rho(0)\ln\rho(0)]-\mathrm{Tr}[\rho(\tau)\ln\rho(\tau)]
    \nonumber\\
    &+\int_{0}^{\tau}dt\sum_{m}\Delta s_{m}\mathrm{Tr}[L_{m}\rho(t)L_{m}^{\dagger}].
    \label{eq:def_EP}
\end{align}
Because 
$\rho(t)=U(t) \rho_{\varnothing}(t) U(t)^\dagger$,
from Eq.~\eqref{eq:main_result_moment} for $ \mathbb{E}[\sum_m \Delta s_m N_m(t)]=\sum_m \Delta s_m \mathrm{Tr}[L_m \rho(t) L_m^\dagger]$, we obtain  the  CIC for the entropy production:
\begin{align}
    \Sigma(\tau)=\Sigma_{\varnothing}(\tau).
    \label{eq:EP_system0}
\end{align}
Equation~\eqref{eq:EP_system0} demonstrates that the entropy production is independent of the presence or absence of coherent contributions in the system.
In the following, we consider an anti-symmetric counting observable
\begin{align}
    J(\tau):=\sum_m c_m N_m(\tau)\;\;\;(c_{m^\prime}=-c_m).
    \label{eq:J_current_def}
\end{align}
$J(\tau)$ is a subset of $N^C(\tau)$ defined in Eq.~\eqref{eq:M_def}, where the additional assumption $c_{m'}=-c_m$ is required. 
Although Eq.~\eqref{eq:main_TUR} was shown via the quantum Fisher information~\cite{Gammelmark:2014:QCRB} 
in Ref.~\cite{Hasegawa:2023:BulkBoundaryBoundNC}, it is difficult to derive a classical-type trade-off relation following a similar procedure for the entropy production
 (see details in the supplementary material \cite{Supp:2025:CIC}).

We introduce the \textit{forward} and \textit{backward} probabilities in $\mathcal{S}_{\varnothing}$ as in the classical dynamics.
Let $ m^{\mathrm{B}}_j:= m_{(K+1-j)^\prime}$, $t_j^{\mathrm{B}}:=\tau - t_{K+1-j}$ for $j\geq 1$ and $t_0^{\mathrm{B}}:=0$. Note that $\{t_j^{\mathrm{B}}\}$ are ordered as $0=t_0^{\mathrm{B}}\le t_1^{\mathrm{B}}< \cdots < t_K^{\mathrm{B}}< \tau$.
The time-reversed trajectory is defined by 
\begin{align}
    \zeta^{\mathrm{B}}_\tau := \left[\left(t^{\mathrm{B}}_1, m^{\mathrm{B}}_1\right), \left(t^{\mathrm{B}}_2, m^{\mathrm{B}}_2\right), \ldots, \left(t^{\mathrm{B}}_K, m^{\mathrm{B}}_K\right);s\right].
    \label{eq:zeta_tau_B_def}
\end{align}
Let $\rho_{\varnothing}(0)=\sum_i q_{i}(0)\ket{i}\bra{i}$ and $\rho_{\varnothing}(\tau)=\sum_{i^\prime} q_{i^\prime}(\tau)\ket{i^\prime}\bra{i^\prime}$ be the spectral decomposition at time $t=0$ and $\tau$, respectively. 
The forward probability $P(i,i^{\prime},\zeta_{\tau})$ is defined as the probability of being $i^\prime$ at time $t=\tau$ from initial state $i$ with probability $q_{i}(0)$ through a trajectory $\zeta_\tau$. Similarly, the backward probability $Q(i,i^{\prime},\zeta_{\tau}^{\mathrm{B}})$ is defined as the probability of being $i$ at time $t=\tau$ from initial state $i^\prime$ with probability $q_{i^\prime}(\tau)$ through a trajectory $\zeta_\tau^\mathrm{B}$. The detailed definitions are shown in the End Matter.
The entropy production in $\mathcal{S}_{\varnothing}$ can be expressed as 
\begin{align}
    \Sigma_{\varnothing}(\tau)=D(P\| Q)&:=\sum_{i,i^\prime} \int\mathcal{D}\zeta_\tau P(i, i^\prime, \zeta_\tau) \ln \frac{P(i, i^\prime, \zeta_\tau)}{Q(i, i^\prime, \zeta_\tau^\mathrm{B})},
    \label{eq:EP_KL_div_main}
\end{align}
where $D(P\|Q)$ denotes the Kullback-Leibler divergence between $P$ and $Q$. The proof of Eq.~\eqref{eq:EP_KL_div_main} is shown in the supplementary material \cite{Supp:2025:CIC}.
Let $\mathbb{E}_{P}[\bullet]$ and $\mathbb{E}_{Q}[\bullet]$ be expectation values with respect to $P$ and $Q$, respectively. We can prove the CIC for expectation values with respect to $P$ and $Q$:
\begin{align}
    \mathbb{E}_{P}[J(\tau)^r]&=\mathbb{E}[J(\tau)^r], \label{eq:forward_moment_main} \\
     \mathbb{E}_{Q}[J(\tau)]&=-\mathbb{E}[J([\tau,2\tau])], \label{eq:back_mean_main}\\
    \mathrm{Var}_{Q}[J(\tau)]&=\mathrm{Var}[J([\tau,2\tau])],\label{eq:back_variance_main}
\end{align}
where $J([t_1,t_2])$ denotes the observable of the trajectory in the interval $[t_1,t_2]$.
 In the system $\mathcal{S}$, Eq.~\eqref{eq:EP_KL_div_main} and Eqs.~\eqref{eq:back_mean_main},~\eqref{eq:back_variance_main} do not simultaneously hold.
The reason and proofs of Eqs.~\eqref{eq:forward_moment_main},~\eqref{eq:back_mean_main}, and~\eqref{eq:back_variance_main} are shown in the End Matter
and the supplementary material \cite{Supp:2025:CIC}.

The lower bound of the Kullback-Leibler divergence is given by the mean and variance of $P$ and $Q$~\cite{Nishiyama:2020:Entropy}. 
By using this lower bound and  the CIC [Eqs.~\eqref{eq:EP_system0},~\eqref{eq:forward_moment_main},~\eqref{eq:back_mean_main}, and~\eqref{eq:back_variance_main}], 
we obtain the TUR in the original system $\mathcal{S}$: 
\begin{align}
    R(\tau):=\gamma(\tau)\frac{\mathrm{Var}[J(\tau)]}{\mathbb{E}[J(\tau)]^2}\geq \mathrm{csch}^2\left[h\left(\frac{\Sigma(\tau)}{2}\right)\right]\geq \frac{2}{e^{\Sigma(\tau)}-1},
    \label{eq:TUR_EP_main}
\end{align}
where $\mathrm{csch}(x)$ is a hyperbolic cosecant, $h(x)$ is the inverse of the function $x\mathrm{tanh}(x)$~\cite{Timpanaro:2019:EFTTUR}, and
$\gamma(\tau)$ is
\begin{align}
    \gamma(\tau):=4\max\left(\frac{\mathrm{Var}[J(\tau/2)]}{\mathrm{Var}[J(\tau)]}, \frac{\mathrm{Var}[J([\tau/2,\tau])]}{\mathrm{Var}[J(\tau)]}\right). 
    \label{eq:def_gamma}
\end{align}
Equation~\eqref{eq:TUR_EP_main} is the second result of this Letter,
whose proof is shown in the supplementary material \cite{Supp:2025:CIC}.
Although the left-hand side of Eq.~\eqref{eq:TUR_EP_main} does not represent the relative variance, the variance divided by the squared mean, 
Eq.~\eqref{eq:TUR_EP_main} is advantageous because its right-hand side only includes the entropy production during the interval $[0,\tau]$. 
Under the steady-state condition, Eq.~\eqref{eq:TUR_EP_main} reduces to
\begin{align}
    \frac{\mathrm{Var}[J(\tau)]}{\mathbb{E}[J(\tau)]^2}\geq \mathrm{csch}^2\left[h\left(\frac{\sigma \tau}{2}\right)\right]\geq \frac{2}{e^{\sigma \tau}-1},
    \label{eq:TUR_EP_ss_main}
\end{align}
where the inequalities involving the middle and right terms are identical to those in Refs.~\cite{Timpanaro:2019:EFTTUR,Hasegawa:2019:FTUR}, respectively. 
When considering coherent jump operators as
considered later
in Eq.~\eqref{eq:Deg_H_def}, non-diagonal terms remain even in the steady-state density operator.  
Nevertheless, Eq.~\eqref{eq:TUR_EP_ss_main} holds, which is a result that cannot be obtained from the conventional classical derivations.

A notable application of Eq.~\eqref{eq:TUR_EP_main} is thermodynamic inference. 
In recent years, researchers have applied TURs to estimate entropy production, as discussed in various studies \cite{Li:2019:EPInference,Manikandan:2019:InferEPPRL,Vu:2020:EPInferPRE,Otsubo:2020:EPInferPRE}. Since Eq.~\eqref{eq:TUR_EP_main} can be regarded as a refined version of the second law of thermodynamics that utilizes the moments of observables, it establishes a lower bound for entropy production based on these moments. Specifically, $\Sigma(\tau)$
is bounded from below by
\begin{align}
    \Sigma(\tau) \geq \frac{2 \operatorname{arcsinh}\left(\frac{1}{\sqrt{R(\tau)}}\right)}{\sqrt{R(\tau)+1}}.
    \label{eq:Sigma_lowerbound_TUR}
\end{align}
The lower bound for the entropy production is expressed by Eq.~\eqref{eq:Sigma_lowerbound_TUR}, which is based solely on the moments of the current $J(\tau)$.
The advantage of Eq.~\eqref{eq:Sigma_lowerbound_TUR} is that it does not include quantum correction terms, which are often challenging to measure experimentally. The entropy production inference using the derived inequality [Eq.~\eqref{eq:Sigma_lowerbound_TUR}] requires only knowledge of the statistics of $J$.

\begin{figure}
\includegraphics[width=1\linewidth]{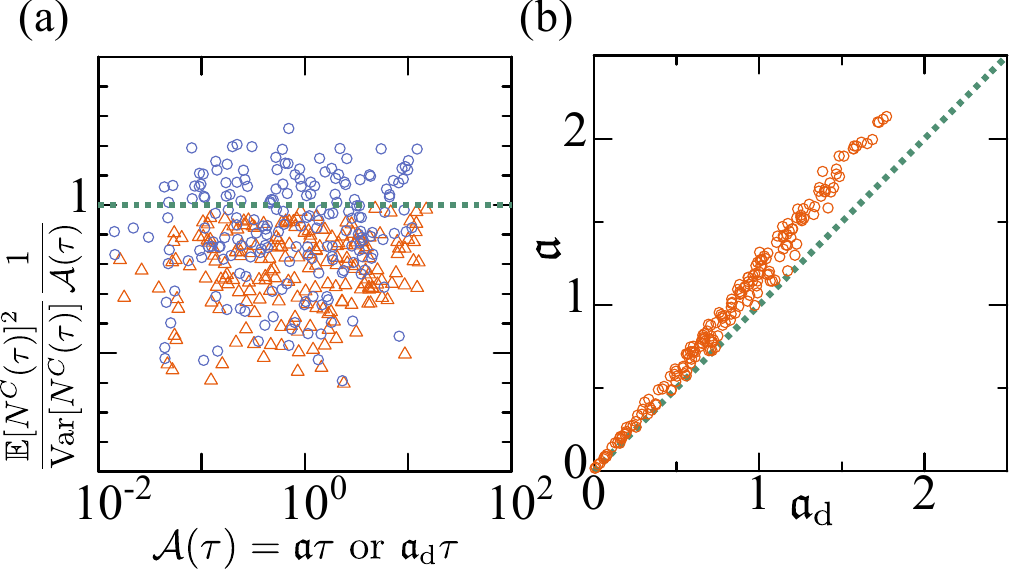}
    \caption{
    Results of numerical simulation of Eq.~\eqref{eq:KUR_steadystate} in the steady-state coherent model.
    (a)
    For each random realization, $\mathbb{E}[N^{C}(\tau)]^{2}/(\mathrm{Var}[N^{C}(\tau)]\mathcal{A}(\tau))$ is plotted against $\mathcal{A}(\tau)$.
    We consider two cases for $\mathcal{A}(\tau)$:
    the dynamical activity $\mathcal{A}(\tau)=\mathfrak{a}\tau$ (red triangle) and the diagonal element $\mathcal{A}(\tau)=\mathfrak{a}_{\mathrm{d}}\tau$ (blue circle). 
    When Eq.~\eqref{eq:KUR_steadystate} holds, all points should be no larger than $1$. 
    (b) 
    Comparison of the dynamical activity $\mathfrak{a}$ and the diagonal element $\mathfrak{a}_\mathrm{d}$. The dotted line indicates the equality case of the dynamical activity and the diagonal element.
    }
    \label{fig:numerical_sim_DA}
\end{figure}

\textit{Example.}---
Let us consider a two-level system with degenerate excited states \cite{Supp:2025:CIC}.
\nocite{Nakajima:2023:SLD,Verstraete:2010:cMPS,Osborne:2010:Holography,Hasegawa:2024:ConcentrationIneqPRL,Valentin:2007:TailProb,Taddei:2013:QSL,Nishiyama:2020:HellingerBound,Helstrom:1976:QuantumEst}
Let $\ket{g}$ be the ground state and $\ket{e_1}$ and $\ket{e_2}$ be degenerate excited states. The Hamiltonian is 
\begin{align}
    H=\omega_{E}(\ket{e_{1}}\bra{e_{1}}+\ket{e_{2}}\bra{e_{2}}),
    \label{eq:Deg_H_def}
\end{align}
where $\omega_E$ is the energy level of the excited states (the ground state energy is assumed to be $0$). 
We consider the following jump operators, 
$L_{1} = \sqrt{\gamma_{1}} \ket{g} \left( \bra{e_{1}} + \bra{e_{2}} \right)$, $L_{2} = \sqrt{\gamma_{2}} \left( \ket{e_{1}} + \ket{e_{2}} \right) \bra{g}$, $L_{3} = \sqrt{\gamma_{3}} \ket{g} \bra{e_{1}}$, and $L_{4} = \sqrt{\gamma_{4}} \ket{g} \bra{e_{2}}$, 
where $\gamma_i$ are the transition rates. 
Non-diagonal elements appear in the steady-state density operator $\rho_\mathrm{ss}$, which is a signature of quantum behavior \cite{Supp:2025:CIC} and distinguishes the dynamics from classical systems.
Since this coherent model still satisfies the condition of Eq.~\eqref{eq:condition},
Eq.~\eqref{eq:KUR_steadystate} holds even in the presence of such coherence.
The dynamical activity $\mathfrak{a}$ in Eq.~\eqref{eq:KUR_steadystate} includes contributions from the diagonal $\mathfrak{a}_\mathrm{d}$ and non-diagonal $\mathfrak{a}_\mathrm{nd}$ elements:
\begin{align}
\mathfrak{a}=\mathrm{Tr}\left[\sum_{m}L_{m}\rho_{\mathrm{ss}}L_{m}^{\dagger}\right]=\mathfrak{a}_{\mathrm{d}}+\mathfrak{a}_{\mathrm{nd}}.
\label{eq:mathfraka_coherent}
\end{align}
Because $\mathfrak{a}_\mathrm{nd}$ is shown to be non-negative in this model, the bound is no larger than the classical counterpart which only includes contributions from the diagonal elements.

We perform a numerical simulation for the coherent model under the steady-state condition. 
In the aforementioned coherent model, the parameters $\gamma_n$ ($n = \{1, 2, 3, 4\}$) are chosen randomly, and the variance, mean of the observable $N^C(\tau)$, and dynamical activity are calculated for each case. In each case, the coefficients $c_m$ and the time $\tau$ are also chosen randomly (for the range of values, see the supplementary material \cite{Supp:2025:CIC}). In Fig.~\ref{fig:numerical_sim_DA}, for each random realization, $\mathbb{E}[N^{C}(\tau)]^{2}/(\mathrm{Var}[N^{C}(\tau)]\mathcal{A}(\tau))$,
which is the ratio between the right and left sides of Eq.~\eqref{eq:KUR_steadystate}, 
is plotted against $\mathcal{A}(\tau)$ for two cases: $\mathcal{A}(\tau)=\mathfrak{a}\tau$ (triangles) and $\mathcal{A}(\tau)=\mathfrak{a}_\mathrm{d}\tau$ (circles). Note that the ratio $\mathbb{E}[N^{C}(\tau)]^{2}/(\mathrm{Var}[N^{C}(\tau)]\mathcal{A}(\tau))$ is no larger than $1$ when Eq.~\eqref{eq:KUR_steadystate} is satisfied. Apparently, all triangles are below $1$, which confirms Eq.~\eqref{eq:KUR_steadystate}.
However, when we replace the dynamical activity with its diagonal contribution (i.e., $\mathcal{A}(\tau) = \mathfrak{a}_\mathrm{d}\tau$), some circles are above $1$, indicating that Eq.~\eqref{eq:KUR_steadystate} with $\mathcal{A}(\tau)$ replaced by the diagonal contribution, is not satisfied. 
This indicates that the system's coherence improves accuracy.
Figure~\ref{fig:numerical_sim_DA}(b) shows a comparison between the dynamical activity $\mathfrak{a}$ and its diagonal contribution $\mathfrak{a}_\mathrm{d}$.  
It can be seen that, for all data points, $\mathfrak{a}$ is greater than $\mathfrak{a}_\mathrm{d}$.

We also perform numerical simulations to investigate the TUR described by Eq.~\eqref{eq:Sigma_lowerbound_TUR},
whose results are shown in the supplementary material \cite{Supp:2025:CIC}. Our results confirm the validity of the bound and demonstrate that the presence of non-diagonal elements enhances the precision of the current $J(\tau)$.

\textit{Conclusion.---}
This Letter introduces the CIC, a method that links the quantities of an open quantum system evolving coherently to those of a corresponding incoherent system without coherent dynamics.
TURs can be readily derived for these incoherent systems owing to their classical-like nature. By using the CIC to map these relations, which hold in the incoherent system, back to the original coherent system, we derived quantum TURs (see Table~\ref{tab:results} for a summary).
Notably, the lower bounds of these relations involve dynamical activity or entropy production without the explicit quantum correction terms present in other existing formulations. 
 This result is particularly significant because it provides bounds that are both structurally simpler and more experimentally accessible, as they do not require the measurement of coherence-dependent quantities. For instance, our entropy production bound [Eq.~\eqref{eq:Sigma_lowerbound_TUR}] offers a practical tool for thermodynamic inference, allowing for the estimation of entropy production based solely on the statistics of an observable current. 
Our framework provides a unified and straightforward approach for establishing trade-off relations in quantum thermodynamics.

\begin{acknowledgments}

This work was supported by JSPS KAKENHI Grant Number JP23K24915.

\end{acknowledgments}

\appendix

\EndMatter

In this End Matter, we show derivations of the derived results.
Their detailed derivations are shown in the supplementary material \cite{Supp:2025:CIC}. 
We show a summary of the obtained uncertainty relations in Table~\ref{tab:results}.

\section{Derivation of Eq.~\eqref{eq:UL_commute1}}

Recall that $U(t)=e^{-iHt}$. Differentiating $\mathcal{F}(t):=U(t)^\dagger L_m U(t)$ with respect to $t$ and using Eq.~\eqref{eq:condition}, we obtain $\dot{\mathcal{F}}(t)=-i\omega_m \mathcal{F}(t)$, and the solution is given by $\mathcal{F}(t)=e^{-i\omega_m t}F(0)=e^{-i\omega_m t}L_m$. Hence, Eq.~\eqref{eq:UL_commute1} follows.

\section{Definition of forward and backward probabilities}

Let $\mathfrak{L}(t):=e^{-\sum_m L_m^{\dagger} L_mt/2}$.
The forward and backward probabilities are defined as
\begin{align}
    &P(i,i^{\prime},\zeta_{\tau}):=\nonumber\\&q_{i}(0)\left|\braket{i^{\prime}|\mathfrak{L}(\tau-t_{K})\mathbb{T}\prod_{j=1}^{K}\left(L_{m_{j}}\mathfrak{L}(t_{j}-t_{j-1})\right)|i}\right|^{2},\label{eq:def_pf_main}\\&Q(i,i^{\prime},\zeta_{\tau}^{\mathrm{B}}):=\nonumber\\&q_{i^{\prime}}(\tau)\left|\braket{i|\mathfrak{L}(\tau-t_{K}^{\mathrm{B}})\mathbb{T}\prod_{j=1}^{K}\left(L_{m_{j}^{\mathrm{B}}}\mathfrak{L}(t_{j}^{\mathrm{B}}-t_{j-1}^{\mathrm{B}})\right)|i^{\prime}}\right|^{2}.\label{eq:def_pb}
\end{align}
Here, $\mathbb{T}$ denotes the time-ordering operator, and we define $\prod_{j=1}^{K}(\bullet):=1$ for $K=0$. 
These probabilities satisfy $\sum_{i, i^\prime} \int\mathcal{D}\zeta_\tau P(i, i^\prime, \zeta_\tau)=\sum_{i, i^\prime} \int\mathcal{D}\zeta_\tau Q(i, i^\prime, \zeta_\tau^\mathrm{B})=1$, where $\int\mathcal{D}\zeta_\tau$ denotes the sum over all trajectories in $[0,\tau]$. 

\section{CIC for Moments of Trajectory Observables}
We begin by establishing a relationship between the expectations of moments of $N(\tau)$ in the two systems $\mathcal{S}$ and $\mathcal{S}_\varnothing$. 
Let $H_{\mathrm{eff}} := H-(i/2) \sum_{m=1}^{N_C} L_m^{\dagger} L_m$ be the effective Hamiltonian in the Lindblad equation.
Because $[H, L_m^\dagger L_m]=0$ holds in $\mathcal{S}$ according to Eq.~\eqref{eq:condition},
$V(t):= e^{-iH_{\mathrm{eff}}t}$, which describes the non-Hermitian time evolution of no-jump dynamics, is decomposed as follows:
\begin{align}
    V(t) = e^{-iH_{\mathrm{eff}}t}=U(t)\mathfrak{L}(t)=\mathfrak{L}(t)U(t).
    \label{eq:U_decomposition}
\end{align}

Consider a purification of the system state $\rho(t)$ as $\ket{\psi(t)} =\sum_{i}\sqrt{p_{i}(t)}\ket{\tilde{i}(t)}\otimes\ket{a_{i}}$, where $\{\ket{a_i}\}$ is an ancillary orthonormal basis. For the incoherent system $\mathcal{S}_\varnothing$ (with $H=0$), a similar purification is $\ket{\psi_\varnothing(t)} =\sum_{i}\sqrt{q_{i}(t)}\ket{i(t)}\otimes\ket{a_{i}}$, and these are related by $\ket{\psi_\varnothing(t)} = U(t)^\dagger \ket{\psi(t)}$, with $U(t) = e^{-iHt}$.

The moment of the trajectory observable $N(\tau)$ in the original (coherent) system $\mathcal{S}$ can be written as:
\begin{align}
    &\mathbb{E}[N(\tau)^r]\nonumber\\
    &=\int\mathcal{D}\zeta_\tau  N(\zeta_\tau)^r \left |
     V(\tau-t_K)  \mathbb{T}\prod_{j=1}^{K} \left(L_{m_j} V(t_{j}- t_{j-1})\right)\ket{\psi(0)}\right|^2.
    \label{eq:moment_S}
\end{align}
For the incoherent system $\mathcal{S}_\varnothing$ ($H=0$),
\begin{align}
    &\mathbb{E}_\varnothing[N(\tau)^r]\nonumber\\
    &=\int\mathcal{D}\zeta_\tau N(\zeta_\tau)^r\left|  \mathfrak{L}(\tau-t_K)  \mathbb{T}\prod_{j=1}^{K} \left( L_{m_j} \mathfrak{L}(t_{j}-t_{j-1})\right)\ket{\psi(0)}\right|^2.
    \label{eq:moment_S0}
\end{align}
Using the commutation relations $L_m U(t) = e^{-i\omega_m t} U(t) L_m$ and $[U(t),\mathfrak{L}(t)] = 0$, the above expressions differ only by an unobservable phase factor,
which indicates the following relation:
\begin{align}
    \mathbb{E}[N(\tau)^r] = \mathbb{E}_\varnothing[N(\tau)^r].
    \label{eq:E_Evarnothing_equiv_EM}
\end{align}
In particular, the dynamical activity $\mathcal{A}(\tau) := \mathbb{E}\bigl[ \sum_m N_m(\tau) \bigr]$ satisfies
\begin{align}
    \mathcal{A}(\tau) = \mathcal{A}_\varnothing(\tau).
    \label{eq:DA_equiv_EM}
\end{align}
Thus, any inequality involving these moments for $\mathcal{S}_\varnothing$ holds for $\mathcal{S}$ as well.

Next, we show the derivation of the bounds for the entropy production.
Since the forward probability $P(i,i^\prime,\zeta_\tau)$, defined in Eq.~\eqref{eq:def_pf_main}, is the probability without coherent dynamics, 
we simply obtain
\begin{align}
    \mathbb{E}_P[N(\tau)^r] = \mathbb{E}_\varnothing[N(\tau)^r].
    \label{eq:P_empty_equivalence_EM}
\end{align}
Moreover, from Eq.~\eqref{eq:E_Evarnothing_equiv_EM}, we have
Eq.~\eqref{eq:forward_moment_main} in the main text. 
The situation with the backward probability $Q(i,i^{\prime},\zeta_{\tau}^{\mathrm{B}})$, defined in Eq.~\eqref{eq:def_pb}, is more complicated.
Recall that $Q(i,i^{\prime},\zeta_{\tau}^{\mathrm{B}})$ represents the probability in the absence of coherent evolution.
Using properties $N^{C}_{m} (\zeta_\tau)=N^{C}_{m^\prime} (\zeta_\tau^{\mathrm{B}})$,
the expectation value with respect to $Q(i,i^{\prime},\zeta_{\tau}^{\mathrm{B}})$ is
\begin{widetext}
\begin{align}
    \mathbb{E}_{Q}[N_{m}^{C}(\tau)]&=\sum_{i,i^{\prime}}\int\mathcal{D}\zeta_{\tau}\,Q(i,i^{\prime},\zeta_{\tau}^{\mathrm{B}})N_{m}^{C}(\zeta_{\tau})\nonumber=\int\mathcal{D}\zeta_{\tau}N_{m^{\prime}}^{C}(\zeta_{\tau})\left|\mathfrak{L}(\tau-t_{K})\prod_{j=1}^{K}\left(L_{m_{j}}\mathfrak{L}(t_{j}-t_{j-1})\right)\ket{\psi_{\varnothing}(\tau)}\right|^{2}\nonumber\\&=\int\mathcal{D}\zeta_{\tau}N_{m^{\prime}}^{C}(\zeta_{\tau})\left|V(\tau-t_{K})\prod_{j=1}^{K}\left(L_{m_{j}}V(t_{j}-t_{j-1})\right)\ket{\psi(\tau)}\right|^{2}.
    \label{eq:mean_Q_pathint_EM}
\end{align}
\end{widetext}
Refer to the supplementary material for details of the derivation \cite{Supp:2025:CIC}.
From the third to the fourth line of Eq.~\eqref{eq:mean_Q_pathint_EM},
we used Eq.~\eqref{eq:UL_commute1} and $[U(t),\mathfrak{L}(t)] = 0$. 
Because the last term in Eq.~\eqref{eq:mean_Q_pathint_EM} is the result of the time evolution from $t=0$ to $\tau$ under the initial condition $\rho(\tau)$, we can identify this term with the time evolution from $t=\tau$ to $2\tau$ with the initial condition $\rho(0)$.
Therefore, we obtain
\begin{align}
     \mathbb{E}_{Q}[N^{C}_{m}(\tau)]=\mathbb{E}_{\varnothing}[N^{C}_{m^\prime}([\tau,2\tau])]=\mathbb{E}[N^{C}_{m^\prime}([\tau,2\tau])].
     \label{eq:EQ_NmC_EM}
\end{align}
Similarly, it follows that 
\begin{align}
    \mathbb{E}_{Q}[N_{m}^{C}(\tau)N_{l}^{C}(\tau)]&=\mathbb{E}_{\varnothing}[N_{m^{\prime}}^{C}([\tau,2\tau])N_{l^{\prime}}^{C}([\tau,2\tau])]\nonumber\\&=\mathbb{E}[N_{m^{\prime}}^{C}([\tau,2\tau])N_{l^{\prime}}^{C}([\tau,2\tau])].
    \label{eq:EQ_NmC_NlC_EM}
\end{align}
By combining these relations with $c_{m^\prime}=-c_{m}$, we obtain Eqs.~\eqref{eq:back_mean_main} and~\eqref{eq:back_variance_main}.

\begin{table*}
\begin{tabular}{cccccc}
\toprule
Bound & Observable & Cost & Dynamics & Assumption & Reference \\
\midrule
$\displaystyle \left(\frac{\sqrt{\mathrm{Var}[N(\tau_{2})]}+\sqrt{\mathrm{Var}[N(\tau_{1})]}}{\mathbb{E}[N(\tau_{2})]-\mathbb{E}[N(\tau_{1})]}\right)^{2}\geq\tan\left[\frac{1}{2}\int_{\tau_{1}}^{\tau_{2}}\frac{\sqrt{\mathcal{A}(t)}}{t}dt\right]^{-2}$ 
& $N(\tau_1),N(\tau_2)$ & $\mathcal{A}(\tau)$ & Transient & Eq.~\eqref{eq:condition} & Eq.~\eqref{eq:main_TUR} \\

$\displaystyle \frac{\mathrm{Var}[N(\tau)]}{\tau^{2}\left(\partial_{\tau}\mathbb{E}[N(\tau)]\right)^{2}}\geq\frac{1}{\mathcal{A}(\tau)}$ 
& $N(\tau)$ & $\mathcal{A}(\tau)$ & Transient & Eq.~\eqref{eq:condition} & Eq.~\eqref{eq:main_TUR_diff} \\

$\displaystyle \frac{\mathrm{Var}[N^{C}(\tau)]}{\mathbb{E}[N^{C}(\tau)]^{2}}\geq\frac{1}{\mathfrak{a}\tau}$ 
& $N^C(\tau)$ & $\mathfrak{a}\tau $ & Steady state & Eq.~\eqref{eq:condition} & Eq.~\eqref{eq:KUR_steadystate} \\

$\displaystyle \frac{\mathbb{E}[|N(\tau)|^{s}]^{r/(s-r)}}{\mathbb{E}[|N(\tau)|^{r}]^{s/(s-r)}}\ge\frac{1}{1-e^{-\mathfrak{a}(0)\tau}}.$ 
& $N(\tau)$ & $\mathfrak{a}(0)\tau $ & Transient & Eq.~\eqref{eq:condition} & Eq.~\concentrationUDAUratio{} \\

$\displaystyle \gamma(\tau)\frac{\mathrm{Var}[J(\tau)]}{\mathbb{E}[J(\tau)]^2}\geq \mathrm{csch}^2\left[h\left(\frac{\Sigma(\tau)}{2}\right)\right]\geq \frac{2}{e^{\Sigma(\tau)}-1}$ 
& $J(\tau)$ & $\Sigma(\tau)$ & Transient & Eqs.~\eqref{eq:condition} \& \eqref{eq:ldb} & Eq.~\eqref{eq:TUR_EP_main} \\

$\displaystyle \frac{\mathrm{Var}[J(\tau)]}{\mathbb{E}[J(\tau)]^2}\geq \mathrm{csch}^2\left[h\left(\frac{\sigma \tau}{2}\right)\right]\geq \frac{2}{e^{\sigma \tau}-1}$ 
& $J(\tau)$ & $\sigma \tau $ & Steady state & Eqs.~\eqref{eq:condition} \& \eqref{eq:ldb} & Eq.~\eqref{eq:TUR_EP_ss_main} \\
\bottomrule
\end{tabular}
\caption{Summary of results. 
TURs for the Lindblad dynamics [Eq.~\eqref{eq:Lindblad_def}].
$N(\tau)$ represents a trajectory observable, $N^C(\tau)$ is a counting observable, and $J(\tau)$ is a current observable. The current observables $J(\tau)$ are a subset of the counting observables $N^C(\tau)$, which themselves are a subset of the trajectory observables $N(\tau)$.
$\mathfrak{a}(t)$ and $\sigma(t)$ are the rates of $\mathcal{A}(t)$ and $\Sigma(t)$, respectively. 
When the system is in a steady state, $\mathcal{A}(\tau) =\mathfrak{a}\tau$ and $\Sigma(\tau) = \sigma \tau$. 
$\mathbb{E}[\bullet]$ and $\mathrm{Var}[\bullet]$ are the expectation value and the variance, respectively. 
$\gamma(\tau)$ is defined in Eq.~\eqref{eq:def_gamma}.
Equation~\concentrationUDAUratio{} is shown in the supplementary material \cite{Supp:2025:CIC}.
$r$ and $s$ should satisfy $0<r<s$.
The ``Observable'' column indicates the target observable. ``Cost'' refers to the thermodynamic cost. The ``Dynamics'' column specifies whether the result applies to transient dynamics or steady-state conditions. 
``Assumption'' outlines the assumptions to derive each result.
}
\label{tab:results}
\end{table*}

\end{document}


\title{Supplementary Material for\\ ``Quantum Thermodynamic Uncertainty Relations without Quantum Corrections:\\ A Coherent-Incoherent Correspondence Approach''}

\author{Tomohiro Nishiyama}
\email{htam0ybboh@gmail.com}
\affiliation{Independent Researcher, Tokyo 206-0003, Japan}

\author{Yoshihiko Hasegawa}
\email{hasegawa@biom.t.u-tokyo.ac.jp}
\affiliation{Department of Information and Communication Engineering, Graduate
School of Information Science and Technology, The University of Tokyo,
Tokyo 113-8656, Japan}

\maketitle
This supplementary material describes the calculations introduced in the main text. The numbers of the equations and the figures are prefixed with S (e.g., Eq.~(S1) or Fig.~S1). Numbers without this prefix (e.g., Eq.~(1) or Fig.~1) refer to items in the main text.

\section{Expression of quantum dynamical activity $\mathcal{B}(\tau)$\label{sec:QDA_def}}
We provide the definition of the quantum dynamical activity $\mathcal{B}(\tau)$ considered in the main text. 
We define the adjoint superoperator $\mathcal{L}^\dagger$ as follows:
\begin{align}
\mathcal{L}^{\dagger}\mathcal{O}:= i\left[H,\mathcal{O}\right]+\sum_{m=1}^{N_{C}}\mathcal{D}^{\dagger}\left[L_{m}\right]\mathcal{O},
\label{eq:mathcalL_dag_def}
\end{align}
where $\mathcal{D}^{\dagger}$ is the adjoint dissipator with $\mathcal{O}$ being an operator:
\begin{align}
\mathcal{D}^{\dagger}[L] \mathcal{O} := L^{\dagger} \mathcal{O} L - \frac{1}{2}\left\{L^{\dagger} L, \mathcal{O}\right\}.
\label{eq:mathcalD_dag_def}
\end{align}
The quantum dynamical activity is defined by \cite{Nakajima:2023:SLD,Nishiyama:2024:ExactQDAPRE}
\begin{align}
\mathcal{B}(\tau) = \mathcal{A}(\tau) + \mathcal{C}(\tau),
\label{eq:mathcalB_def}
\end{align}
with
\begin{align}
\mathcal{C}(\tau):= 8\int_{0}^{\tau}ds_{1}\int_{0}^{s_{1}}ds_{2}\mathrm{Re}\left[\mathrm{Tr}\left\{ H_{\mathrm{eff}}^{\dagger}\check{H}\left(s_{1}-s_{2}\right)\rho\left(s_{2}\right)\right\} \right]-4\left(\int_{0}^{\tau}ds\mathrm{Tr}\left[H\rho(s)\right]\right)^{2},
\label{eq:mathcalC_def}
\end{align}
where $\check{H}(t) := e^{\mathcal{L}^{\dagger} t} H$ represents the Hamiltonian $H$ in the Heisenberg picture.
Please note that the expression in Eq.~\eqref{eq:mathcalC_def} follows Ref.~\cite{Nishiyama:2024:ExactQDAPRE}.  
A different form of this quantity can be found in Ref.~\cite{Nakajima:2023:SLD}.

\section{Scaled continuous matrix product state}
Let $\phi_m(t)$ be a field operator that satisfies the canonical commutation relation $[\phi_{m}(t),\phi_{n}^{\dagger}(t^{\prime})]=\delta_{mn}\delta(t-t^{\prime})$, and let 
$|\mathrm{vac}\rangle$ be the vacuum state. 
The continuous matrix product state $\ket{\Phi(\tau)}$ can be represented as \cite{Verstraete:2010:cMPS,Osborne:2010:Holography}:
\begin{align}
    \ket{\Phi(\tau)}=\mathcal{U}(\tau)\ket{\psi(0)}\otimes\ket{\mathrm{vac}},
    \label{eq:cMPS_def}
\end{align}
where $\mathcal{U}(\tau)$ is the operator defined by
\begin{align}
    \mathcal{U}(\tau)=\mathbb{T}e^{-i\int_{0}^{\tau}dt\,(H_{\mathrm{eff}}\otimes\mathbb{I}_{F}+\sum_{m}iL_{m}\otimes\phi_{m}^{\dagger}(t))}.
    \label{eq:mathcalV_def}
\end{align}
Here, $\mathbb{I}_F$ is the identity operator in the field.
The density operator at $\tau$ is given by $\rho(\tau)=\mathrm{Tr}_F[\ket{\Phi(\tau)}\bra{\Phi(\tau)}]$, where $\mathrm{Tr}_F[\bullet]$ denotes the trace with respect to the field. By applying $\phi_m^{\dagger}(t)$ to the vacuum state, the continuous matrix product state (cMPS) captures all the information of the continuous measurement by generating particles. Let 
\begin{align}
    \int\mathcal{D}\zeta_\tau \bullet:=\sum_{K=0}^{\tau/dt} \sum_{m_1, m_2, \cdots, m_K} \int_0^\tau dt_K\int_0^{t_{K}} dt_{K-1} \cdots \int_0^{t_2} dt_1 \bullet
    \label{eq:def_path_integral}
\end{align}
be a sum over all trajectories in $[0,\tau]$, where we define $\int_0^\tau dt_K\int_0^{t_{K}} dt_{K-1} \cdots \int_0^{t_2} dt_1:=1$ for $K=0$.
The cMPS can be expanded as
\begin{align}
    \ket{\Phi(\tau)}=\int \mathcal{D}\zeta_\tau  V(\tau-t_K)  \prod_{j=1}^{K} \left(\phi_{m_j}^\dagger(t_{j})L_{m_j} V(t_{j}- t_{j-1})\right)\ket{\Phi(0)}.
\label{eq:cMPS_expand}
\end{align}
To consider the fidelity at different times, we define the scaled continuous matrix product state \cite{Hasegawa:2023:BulkBoundaryBoundNC}:
\begin{align}
    \ket{\Psi(\tau)}:=\mathcal{V}(\theta)\ket{\Phi(0)},
    \label{eq:scaled_cMPS}
\end{align}
where $\theta := \tau/ t$ is the scale parameter and
\begin{align}
    \mathcal{V}(\theta):=\mathbb{T}e^{\int_{0}^{t}ds\,(-i\theta H_{\mathrm{eff}}\otimes\mathbb{I}_{\mathrm{F}}+\sqrt{\theta}\sum_{m}L_{m}\otimes\phi_{m}^{\dagger}(s))}.
    \label{eq:mathcalV_theta_def}
\end{align}
As in Eq.~\eqref{eq:cMPS_expand}, the scaled continuous matrix product state can be expanded as 
\begin{align}
    \ket{\Psi(\tau)}=\int \mathcal{D} \zeta_t \ket{\Psi(\zeta_t,\theta)},
    \label{eq:scaled_cMPS_expand}
\end{align}
where 
\begin{align}
    \ket{\Psi(\zeta_t,\theta)}:= \theta^{\frac{K}{2}}V(\tau-\theta s_K)  \prod_{j=1}^{K} \left(\phi_{m_j}^\dagger(s_{j})L_{m_j} V(\theta(s_{j}- s_{j-1}))\right)\ket{\Phi(0)}
    \label{eq:def_scaled_cMPS_zeta}
\end{align}
for $0=s_0 < s_1<\cdots< s_K \le t$. Note that the sum in Eq.~\eqref{eq:scaled_cMPS_expand} is taken over trajectories $\zeta_t$ in $[0,t]$ for fixed $t$.
By using this representation, 
we can evaluate the fidelity $\left|\braket{\Psi(0)|\Psi(\tau)}\right|$ because the integration range in Eq.~\eqref{eq:scaled_cMPS_expand} does not depend on $\tau$. 

\section{TUR for dynamical activity}
\subsection{Proof of CIC for moments of trajectory observables \label{seq:moment_equivalence}}
We prove the CIC for the moments of trajectory observables [Eqs.~\mainUresultUmoment{} and~\mainUresultUmomentUabs{}].
For a spectral decomposition $\rho(t)=\sum_i p_{i}(t)\ket{\tilde{i}(t)}\bra{\tilde{i}(t)}$, let  
\begin{align}
    \ket{\psi(t)}:=\sum_{i}\sqrt{p_{i}(t)}\ket{\tilde{i}(t)}\otimes\ket{a_{i}}
    \label{eq:def_state_ancilla}
\end{align}
be the purified state vector and let $\{\ket{a_{i}}\}$ be an orthonormal basis in the ancilla. 
The density operator is given by $\rho(t)=\mathrm{Tr}_A [\ket{\psi(t)}\bra{\psi(t)}]$, where $\mathrm{Tr}_A[\bullet]$ denotes the trace with respect to the ancilla. Similarly, for a spectral decomposition $\rho_{\varnothing}(t)=\sum_i q_{i}(t)\ket{i(t)}\bra{i(t)}$, let
\begin{align}
    \ket{\psi_{\varnothing}(t)}:=\sum_{i}\sqrt{q_{i}(t)}\ket{i(t)}\otimes\ket{a_{i}}.
    \label{eq:def_state_ancilla_0}
\end{align}
Recalling $\rho_{\varnothing}(t)=U(t)^\dagger \rho(t) U(t)$, we have 
\begin{align}
    \ket{\psi_{\varnothing}(t)}=U(t)^\dagger\ket{\psi(t)}
    \label{eq:psi_unitary}
\end{align}
with $\ket{\psi_{\varnothing}(0)}=\ket{\psi(0)}$.

From Eq.~\eqref{eq:psi_unitary}, and writing the dependence of $N(\tau)$ on the trajectory $\zeta_\tau$ explicitly, the moments $\mathbb{E}[N(\tau)^r]$ and $\mathbb{E}_{\varnothing}[N(\tau)^r]$ are given by
\begin{align}
    \mathbb{E}[N(\tau)^r]&=\int\mathcal{D}\zeta_\tau  N(\zeta_\tau)^r \left |
     V(\tau-t_K)  \mathbb{T}\prod_{j=1}^{K} \left(L_{m_j} V(t_{j}- t_{j-1})\right)\ket{\psi(0)}\right|^2,  \label{eq:moment_S}\\
    \mathbb{E}_{\varnothing}[N(\tau)^r]&=\int\mathcal{D}\zeta_\tau N(\zeta_\tau)^r\left|  \mathfrak{L}(\tau-t_K)  \mathbb{T}\prod_{j=1}^{K} \left( L_{m_j} \mathfrak{L}(t_{j}-t_{j-1})\right)\ket{\psi(0)}\right|^2,
    \label{eq:moment_S0}
\end{align}
where $\mathbb{T}$ denotes the time-ordering operator and we define $\prod_{j=1}^{K=0}(\bullet):=1$ and $t_0:=0$. In the following, we drop $\mathbb{T}$ 
when it is apparent from the context.
From Eq.~\UUdecomposition{}, we obtain
\begin{align}
    &\mathbb{E}[N(\tau)^r]=\int\mathcal{D}\zeta_\tau N(\zeta_\tau)^r \left| U(\tau-t_K)\mathfrak{L}(\tau-t_K)  \prod_{j=1}^{K} \left(L_{m_j} U(t_{j}-t_{j-1})\mathfrak{L}(t_{j}-t_{j-1})\right)\ket{\psi(0)}\right|^2.
    \label{eq:moment_S_2}
\end{align}
Applying Eq.~\ULUcommuteI{} and $[U(t),\mathfrak{L}(t)]=0$ repeatedly, and combining with Eq.~\momentUSO{}, we obtain 
\begin{align}
    &\mathbb{E}[N(\tau)^r]=\int\mathcal{D}\zeta_\tau N(\zeta_\tau)^r\left|e^{if(\zeta_\tau)} U(\tau)\mathfrak{L}(\tau-t_K)  \prod_{j=1}^{K} \left(L_{m_j} \mathfrak{L}(t_{j}-t_{j-1})\right)\ket{\psi(0)}\right|^2=\mathbb{E}_{\varnothing}[N(\tau)^r],
    \label{eq:S0_relation}
\end{align}
where
$f(\zeta_\tau):=-\sum_{j=1}^K (t_j-t_{j-1})\sum_{i=j}^K \omega_{m_i}$
and we use the unitarity of $U(t)$ in the second equality. Eq.~\eqref{eq:S0_relation} completes the proof of Eq.~\mainUresultUmoment{}. Similarly, we can prove Eq.~\mainUresultUmomentUabs{}.

\subsection{TUR for moments of arbitrary order}
We will derive two types of thermodynamic uncertainty relations for moments of arbitrary order.
We assume that the trajectory observable satisfies the following condition:
\begin{align}
    N(\zeta_{0}) = 0,
    \label{eq:Ovarnothing_def}
\end{align}
where $\zeta_{0}$ denotes the trajectory with no jumps.
First, we generalize Eq.~\mainUTUR{} for moments of arbitrary order.
By applying the result in Ref.~\cite{hasegawa2024thermodynamic} for the system $\mathcal{S}_{\varnothing}$ and using $\mathcal{B}_{\varnothing}(t)=\mathcal{A}_{\varnothing}(t)$ from Eq.~\eqref{eq:mathcalB_def} with $H=0$, we obtain 
\begin{align}
    \frac{\mathbb{E}_{\varnothing}[|N(\tau)|^{s}]^{r/(s-r)}}{\mathbb{E}_{\varnothing}[|N(\tau)|^{r}]^{s/(s-r)}}&\ge\sin\left[\frac{1}{2}\int_{0}^{\tau}\frac{\sqrt{\mathcal{A}_{\varnothing}(t)}}{t}dt\right]^{-2},
    \label{eq:concentration_DA_S0}
\end{align}
where $0 < r < s$.
The CIC [Eqs.~\mainUresultUmomentUabs{} and~\DAUSO{}] yields
\begin{align}
    \frac{\mathbb{E}[|N(\tau)|^{s}]^{r/(s-r)}}{\mathbb{E}[|N(\tau)|^{r}]^{s/(s-r)}}&\ge\sin\left[\frac{1}{2}\int_{0}^{\tau}\frac{\sqrt{\mathcal{A}(t)}}{t}dt\right]^{-2},
    \label{eq:concentration_DA}
\end{align}
where we assume that $(1/2)\int_{\tau_1}^{\tau_2}\sqrt{\mathcal{A}(t)}/t\,dt\le\pi/2$.
For $r=1$ and $s=2$, Eq.~\eqref{eq:concentration_DA} recovers the thermodynamic uncertainty relation Eq.~\mainUTUR{} for $\tau_1=0$:
\begin{align}
    \frac{\mathrm{Var}[N(\tau)]}{\mathbb{E}[N(\tau)]^{2}}\geq \frac{\mathrm{Var}[|N(\tau)|]}{\mathbb{E}[|N(\tau)|]^{2}}&\ge\tan\left[\frac{1}{2}\int_{0}^{\tau}\frac{\sqrt{\mathcal{A}(t)}}{t}dt\right]^{-2}.\label{eq:TUR1}
\end{align}
Here, the first inequality in Eq.~\eqref{eq:TUR1} follows from the inequalities $\mathbb{E}[|N(\tau)|]\ge |\mathbb{E}[N(\tau)]|$ and $\mathrm{Var}[|N(\tau)|]\le\mathrm{Var}[N(\tau)]$.

We next derive another type of thermodynamic uncertainty relation. When $[H, \sum_m L_m^\dagger L_m]=0$, from the results in Ref.~\cite{Nishiyama:2024:NonHermiteQSLPRA}, we obtain
\begin{align}
    \left|\braket{\Psi(0)|\Psi(\tau)}\right|\geq e^{-\frac{1}{2}\mathfrak{a}(0)\tau}-\tau\left(\mathbb{E}[H](0) -E_{g}\right),
    \label{eq:fidelity_Levitin}
\end{align}
where $E_{g}$ denotes the minimum eigenvalue of $H$ (the ground state energy).
From Eqs.~\eqref{eq:scaled_cMPS_expand} and $\ket{\psi_{\varnothing}(0)}=\ket{\psi(0)}$, the probability of no jumps $\mathfrak{p}_{\varnothing}(\tau)$ is
\begin{align}
    \mathfrak{p}_{\varnothing}(\tau)=\braket{\psi(0)|\mathfrak{L}(\tau)^{\dagger}\mathfrak{L}(\tau)|\psi(0)},
    \label{eq:pfrak_def}
\end{align}
which satisfies the following inequality:
\begin{align}
    \left|\braket{\Psi_{\varnothing}(0)|\Psi_{\varnothing}(\tau)}\right|^{2}=\left|\braket{\psi(0)|\mathfrak{L}(\tau)|\psi(0)}\right|^{2}
    \le\left|\braket{\psi(0)|\mathfrak{L}(\tau)^{\dagger}\mathfrak{L}(\tau)|\psi(0)}\right|=\mathfrak{p}_{\varnothing}(\tau).
    \label{eq:fidelity_frakp}
\end{align}
For the first inequality, we used the Cauchy-Schwarz inequality. 
From Eq.~\eqref{eq:fidelity_Levitin} for the system $\mathcal{S}_{\varnothing}$ and Eq.~\eqref{eq:fidelity_frakp}, it follows that
\begin{align}
     \mathfrak{p}_{\varnothing}(\tau) \ge e^{-\mathfrak{a}_{\varnothing}(0)\tau}.
     \label{eq:prob_mean}
\end{align}
Let us relate the probability $\mathfrak{p}_{\varnothing}(\tau)$ with the probability of observing no jumps within the time interval $[0, \tau]$, denoted as $P_{\varnothing}(N(\tau) = 0)$. $\mathfrak{p}_\varnothing(\tau)=0$ implies $N(\tau)=0$. However, the converse is not always true. Consequently, we have the following inequality:
\begin{align}
    P_{\varnothing}(N(\tau) = 0) \ge \mathfrak{p}_{\varnothing}(\tau)\ge e^{-\mathfrak{a}_{\varnothing}(0)\tau}.
    \label{eq:P_mathfrakp_ineq}
\end{align}
Reference~\cite{hasegawa2024thermodynamic} used the Petrov inequality (Eq.~(9) in \cite{Valentin:2007:TailProb}) to link the probability $P(N(\tau)=0)$ to the moment of $|N(\tau)|$.
The Petrov inequality states
\begin{align}
    P(|X| > b) \ge \frac{(\mathbb{E}[|X|^r] - b^r)^{s/(s-r)}}{\mathbb{E}[|X|^s]^{r/(s-r)}},
    \label{eq:Paley_Zygmund_ineq}
\end{align}
where $0 < r < s$, $b \ge 0$, and the condition $b^r \le \mathbb{E}[|X|^r]$ must hold. This inequality provides a lower bound on the tail probability of $|X|$ in terms of its moments. By combining the Petrov inequality for $b = 0$ with Eq.~\eqref{eq:P_mathfrakp_ineq} and using the CIC [Eqs.~\mainUresultUmomentUabs{} and~\DAUratioUSO{}], we obtain 
\begin{align}
\frac{\mathbb{E}[|N(\tau)|^{s}]^{r/(s-r)}}{\mathbb{E}[|N(\tau)|^{r}]^{s/(s-r)}}&\ge\frac{1}{1-e^{-\mathfrak{a}(0)\tau}}. \label{eq:concentration_DA_ratio}
\end{align}
As in Eq.~\eqref{eq:TUR1}, it follows that
\begin{align}
    \frac{\mathrm{Var}[N(\tau)]}{\mathbb{E}[N(\tau)]^{2}}\geq \frac{\mathrm{Var}[|N(\tau)|]}{\mathbb{E}[|N(\tau)|]^{2}}&\ge\frac{1}{e^{\mathfrak{a}(0)\tau}-1}.
    \label{eq:TUR1_classical_stronger}
\end{align} 
A notable advantage of Eq.~\eqref{eq:concentration_DA_ratio} over Eq.~\eqref{eq:concentration_DA} is that it
holds for any $\tau> 0$. In the steady-state condition, it was shown that Eq.~\eqref{eq:concentration_DA_ratio} is tighter than Eq.~\eqref{eq:concentration_DA} in Ref.~\cite{hasegawa2024thermodynamic}.

\subsection{Proof of Eq.~\mainUTUR{} in $\mathcal{S}_{\varnothing}$}
We prove Eq.~\mainUTUR{} for the system $\mathcal{S}_{\varnothing}$.
Consider the auxiliary dynamics with keeping $H=0$ and modified jump operator as in Ref.~\cite{Hasegawa:2023:BulkBoundaryBoundNC}:
\begin{align}
    L_{m, \eta}&:=\sqrt{1+\eta}L_m.
    \label{eq:auxiliary_rescale}
\end{align}
Let $I_{\varnothing,\eta}(\tau)$ be the quantum Fisher information in the system $\mathcal{S}_{\varnothing}$ with respect to the parameter $\eta$: 
\begin{align}
    I_{\varnothing,\eta}(\tau):=4\left[\braket{\partial_\eta \Psi_{\varnothing,\eta}(\tau)|\partial_\eta \Psi_{\varnothing,\eta}(\tau)}-\left|\braket{\partial_\eta \Psi_{\varnothing,\eta}(\tau)| \Psi_{\varnothing,\eta}(\tau)}\right|^2\right].
    \label{eq:I_varnothing_eta}
\end{align}
and $\ket{\partial_\eta \Psi_{\varnothing,\eta}(\tau)}:=\partial_\eta \ket{\Psi_{\varnothing,\eta}(\tau)}$. From Eqs.~\eqref{eq:scaled_cMPS_expand} and~\eqref{eq:def_scaled_cMPS_zeta}, the auxiliary dynamics Eq.~\eqref{eq:auxiliary_rescale} converts $\theta$ to $(1+\eta)\theta$ and it follows that $\ket{\Psi_{\varnothing,\eta}(\tau)}=\ket{\Psi_{\varnothing}((1+\eta)\tau)}$.
The upper bound for the fidelity is given by the quantum Fisher information \cite{Taddei:2013:QSL}: 
\begin{align}
    \frac{1}{2}\int_{\eta_1}^{\eta_2} d\eta \sqrt{I_{\varnothing, \eta}(\tau)} \geq \mathcal{L}_D(\ket{\Psi_{\varnothing}((1+\eta_1)\tau)}, \ket{\Psi_{\varnothing}((1+\eta_2)\tau)}).
    \label{eq:QSL_open_quantum_eta}
\end{align}
Here, $\mathcal{L}_D$ is the Bures angle, defined as follows:
\begin{align}
    \mathcal{L}_D(\ket{\Psi(\tau_1)}, \ket{\Psi(\tau_2)})&:= \mathrm{arccos}\left|\braket{\Psi(\tau_2)|\Psi(\tau_1)}\right|.
    \label{eq:Bures_angle_def}
\end{align}
From the result in~\cite{Hasegawa:2023:BulkBoundaryBoundNC}, $I_{\varnothing, \eta}(\tau)$ is given by the quantum dynamical activity $\mathcal{B}_{\varnothing}((1+\eta)\tau) / (1+\eta)^2$. From Eq.~\eqref{eq:mathcalB_def}, $\mathcal{B}_{\varnothing}(\tau)$ reduces to $\mathcal{A}_{\varnothing}(\tau)$ because $H=0$, it follows that
\begin{align}
    \left|\braket{\Psi_{\varnothing}(\tau_2)|\Psi_{\varnothing}(\tau_1)}\right|\ge \cos\left[\frac{1}{2}\int_{\tau_1}^{\tau_2} ds \frac{\sqrt{\mathcal{A}_{\varnothing}(s)}}{s}\right],
    \label{eq:QSL_open_quantum_time}
\end{align}
where $\tau_1:=(1+\eta_1)\tau=(1+\eta_1)\theta t$ and $\tau_2:=(1+\eta_2)\tau=(1+\eta_2)\theta t$.
For $\theta_1:=(1+\eta_1)\theta$ and $\theta_2:=(1+\eta_2)\theta$, let $p(\zeta_t):= 
\braket{\Psi_{\varnothing}(\zeta_t, \theta_1)|\Psi_{\varnothing}(\zeta_t, \theta_1)}$ and $q(\zeta_t):=\braket{\Psi_{\varnothing}(\zeta_t, \theta_2)|\Psi_{\varnothing}(\zeta_t, \theta_2)}$.
Since $\ket{\Psi_{\varnothing}(\tau)}$ is expanded as $\ket{\Psi_{\varnothing}(\tau)}=\int\mathcal{D}\zeta_t\ket{\Psi_{\varnothing}(\zeta_t,\theta)}$ as in Eq.~\eqref{eq:scaled_cMPS_expand}, 
we obtain 
\begin{align}   
    \left|\braket{\Psi_{\varnothing}(\tau_2)|\Psi_{\varnothing}(\tau_1)}\right|= \left|\int\mathcal{D}\zeta_t  \braket{\Psi_{\varnothing}(\zeta_t, \theta_2)|\Psi_{\varnothing}(\zeta_t, \theta_1)}\right|\le \int\mathcal{D}\zeta_t \sqrt{p(\zeta_t)q(\zeta_t)}=1-H^2(p,q),
    \label{eq:fidelity_hellinger_upper_bound}
\end{align}
where $H^2(p,q):= \sum_z (\sqrt{p(z)}-\sqrt{q(z)})^2 /2 = 1-\sum_z \sqrt{p(z)q(z)}$ is the squared Hellinger distance. Let $\mu_p:= \int\mathcal{D}\zeta_t p(\zeta_t) N(\zeta_t) $ and $\sigma_p^2:= \int\mathcal{D}\zeta_t p(\zeta_t) N(\zeta_t)^2 - \mu_p^2$ be the expectation value and the variance, respectively. We analogously define $\mu_q$ and $\sigma_q$. 
Given expectation values and variances, the squared Hellinger distance is lower-bounded by~\cite{Nishiyama:2020:HellingerBound} 
\begin{align}
    H^2(p,q)\geq 1-\left[\left(\frac{\mu_p-\mu_q}{\sigma_p+\sigma_q}\right)^2+1\right]^{-\frac{1}{2}}.
    \label{eq:Hellinger_lower_bound}
\end{align}
From Eqs.~\eqref{eq:def_path_integral},~\eqref{eq:def_scaled_cMPS_zeta} and~\momentUSO{}, we obtain
\begin{align}
    &\int\mathcal{D}\zeta_t N(\zeta_t)^r p(\zeta_t)\nonumber\\
    &=\sum_{K=0}^{\tau_1 /(\theta_1 ds)}\sum_{m_1, m_2, \cdots, m_K} \int_0^t \theta_1 ds_K\int_0^{s_{K}} \theta_1 ds_{K-1} \cdots \int_0^{s_2} \theta_1 ds_1 N(\zeta_t)^r \left| \mathfrak{L}(\tau-\theta_1 s_K)  \prod_{j=1}^{K} \left(L_{m_j} \mathfrak{L}(\theta_1(s_{j}-s_{j-1})\right)\ket{\psi(0)}\right|^2\nonumber\\
    &= \mathbb{E}_{\varnothing}[ N(\tau_1)^r],
    \label{eq:path_integral_Dzetat}
\end{align}
where we use $\tau_1=\theta_1 t$.
Similarly, we obtain $\int\mathcal{D}\zeta_t N(\zeta_t)^r q(\zeta_t)=\mathbb{E}_{\varnothing}[ N(\tau_2)^r]$. Combining these relations with Eqs.~\eqref{eq:QSL_open_quantum_time}, ~\eqref{eq:fidelity_hellinger_upper_bound} and~\eqref{eq:Hellinger_lower_bound}, we obtain 
\begin{align}
\left(\frac{\sqrt{\mathrm{Var}_{\varnothing}[N(\tau_{2})]}+\sqrt{\mathrm{Var}_{\varnothing}[N(\tau_{1})]}}{\mathbb{E}_{\varnothing}[N(\tau_{2})]-\mathbb{E}_{\varnothing}[N(\tau_{1})]}\right)^{2}\geq\tan\left[\frac{1}{2}\int_{\tau_{1}}^{\tau_{2}}\frac{\sqrt{\mathcal{A}_{\varnothing}(s)}}{s}ds\right]^{-2}.
    \label{eq:TUR_0}
\end{align}

\section{TUR for entropy production}
\subsection{Proof of Eq.~\EPUKLUdivUmain{} \label{sec:EP_KL_div}} 
Let $\ket{\zeta_\tau,i}:=\mathfrak{L}(\tau-t_K)  \prod_{j=1}^{K} \left(L_{m_j} \mathfrak{L}(t_{j}-t_{j-1})\right)\ket{i}$. 
From the definition in Eq.~\defUpfUmain{}, the forward probability can be written as 
\begin{align}
    P(i, i^\prime, \zeta_\tau)=q_i(0)|\braket{i^\prime|\zeta_\tau,i}|^2, 
    \label{eq:def_pf}
\end{align}    
which satisfies $\sum_{i, i^\prime} \int\mathcal{D}\zeta_\tau P(i, i^\prime, \zeta_\tau)=\sum_{i}q_i(0)\int\mathcal{D}\zeta_\tau \braket{\zeta_\tau,i|\left(\sum_{i^\prime}\ket{i^\prime}\bra{i^\prime}\right) |\zeta_\tau,i}=\sum_{i}q_i(0)\int\mathcal{D}\zeta_\tau \braket{\zeta_\tau,i|\zeta_\tau,i}=1$. Here we use $\sum_{i^\prime}\ket{i^\prime}\bra{i^\prime}=\mathbb{I}$. Similarly, letting $\ket{\zeta_\tau^{\mathrm{B}},i^\prime}:=\mathfrak{L}(\tau-t^{\mathrm{B}}_K)  \prod_{j=1}^{K} \left(L_{m_j^{\mathrm{B}}} \mathfrak{L}(t^{\mathrm{B}}_{j}-t^{\mathrm{B}}_{j-1})\right)\ket{i^\prime}$, the backward probability in Eq.~\defUpb{} can be written as 
\begin{align}
    Q(i, i^\prime, \zeta_\tau^\mathrm{B})=q_{i^\prime}(\tau)|\braket{i |\zeta_\tau^{\mathrm{B}},i^\prime}|^2,
    \label{eq:Q_i_iprime_zeta}
\end{align}
which satisfies $\sum_{i, i^\prime} \int\mathcal{D}\zeta_\tau Q(i, i^\prime, \zeta_\tau^\mathrm{B})=\sum_{i^\prime} q_{i^\prime}(\tau)\int\mathcal{D}\zeta_\tau\braket{\zeta^{\mathrm{B}}_\tau,i^\prime|\zeta^{\mathrm{B}}_\tau,i^\prime}=\sum_{i^\prime} q_{i^\prime}(\tau)\int\mathcal{D}\zeta_\tau\braket{\zeta_\tau,i^\prime|\zeta_\tau,i^\prime}=1$. Note that the sum over all trajectories for $\zeta_\tau$ is equal to the sum over all trajectories for $\zeta^{\mathrm{B}}_\tau$ because $\zeta_\tau$ and $\zeta^{\mathrm{B}}_\tau$ are in one-to-one correspondence. 
By using $\mathfrak{L}(t)^\dagger = \mathfrak{L}(t)$, the definition of $\zeta^{\mathrm{B}}_\tau$, and Eq.~\ldb{}, we can write Eq.~\defUpb{} as 
\begin{align}
    Q(i, i^\prime, \zeta_\tau^\mathrm{B})&=q_{i^\prime}(\tau)\left| \braket{i| \overline{\mathbb{T}}\prod_{j=1}^{K} \left(\mathfrak{L}(t_{K+1-j}-t_{K-j})L_{m_{{(K+1-j)}^\prime}} \right)\mathfrak{L}(\tau-t_K)|i^\prime}\right|^2 \nonumber\\
    &=q_{i^\prime}(\tau)\left| \braket{i|\overline{\mathbb{T}} \prod_{j=1}^{K}\left( \mathfrak{L}(t_{j}-t_{j-1})L_{m^\prime_j}\right) \mathfrak{L}(\tau-t_K)|i^\prime}\right|^2 \nonumber \\
    &=e^{-\sum_{j=1}^K \Delta s_{m_j}}q_{i^\prime}(\tau)\left| \braket{i^\prime| \mathfrak{L}(\tau-t_K)  \mathbb{T} \prod_{j=1}^{K}\left( L_{m_j} \mathfrak{L}(t_{j}-t_{j-1})\right)|i}\right|^2=e^{-\sum_{j=1}^K \Delta s_{m_j}}\frac{q_{i^\prime}(\tau)}{q_i(0)}  P(i, i^\prime, \zeta_\tau),
    \label{eq:pb_pf_relation} 
\end{align}
where $\overline{\mathbb{T}}$ denotes the anti-time-ordering operator.
Let $D(p\| q):=\sum_z p(z)\ln (p(z)/q(z))$ be the Kullback-Leibler divergence.
From Eqs.~\eqref{eq:def_pf} and~\eqref{eq:pb_pf_relation}, we can associate the Kullback-Leibler divergence between $P$ and $Q$ with the entropy production:
\begin{align}
    D(P\| Q)&:=\sum_{i,i^\prime} \int\mathcal{D}\zeta_\tau P(i, i^\prime, \zeta_\tau) \ln \frac{P(i, i^\prime, \zeta_\tau)}{Q(i, i^\prime, \zeta_\tau^\mathrm{B})}=\sum_{i,i^\prime} \int\mathcal{D}\zeta_\tau P(i, i^\prime, \zeta_\tau)\left[\ln q_i(0)-\ln q_{i^\prime}(\tau) +\sum_{j=1}^K \Delta s_{m_j}\right ]\nonumber\\
    &=\sum_{i,i^\prime} \int\mathcal{D}\zeta_\tau P(i, i^\prime, \zeta_\tau)\left[\ln q_i(0)-\ln q_{i^\prime}(\tau) \right ] + \int\mathcal{D}\zeta_\tau \left(\sum_{j=1}^K \Delta s_{m_j} \right)\left|  \mathfrak{L}(\tau-t_K)  \mathbb{T}\prod_{j=1}^{K} \left( L_{m_j} \mathfrak{L}(t_{j}-t_{j-1})\right)\ket{\psi_{\varnothing}(0)}\right|^2\nonumber\\
    &=\sum_i q_{i}(0)\ln q_{i}(0)-\sum_{i^\prime} q_{i^\prime}(\tau)\ln q_{i^\prime}(\tau)+\int_{0}^{\tau}ds\sum_{m}\Delta s_{m}\mathrm{Tr}[L_{m}\rho_{\varnothing}(s)L_{m}^{\dagger}]=\Sigma_{\varnothing}(\tau).
    \label{eq:EP_KLdiv}
\end{align}
Here, we use $\sum_{i^\prime} \int\mathcal{D}\zeta_\tau P(i, i^\prime, \zeta_\tau)=q_i(0)\int\mathcal{D}\zeta_\tau \braket{\zeta_\tau,i|\zeta_\tau, i}=q_i(0)$, $\sum_{i}q_i(0) \int\mathcal{D}\zeta_\tau \ket{\zeta_\tau,i}\bra{\zeta_\tau, i}=\rho_{\varnothing}(\tau)$ and $\braket{i^\prime|\rho_{\varnothing}(\tau)|i^\prime}=q_{i^\prime}(\tau)$. 
From the definition of entropy production [Eq.~\defUEP{}], Eq.~\eqref{eq:EP_KLdiv} completes the proof of Eq.~\EPUKLUdivUmain{}.

\subsection{Proof of CIC for forward and backward probabilities}
We prove the CIC for anti-symmetric counting observables with respect to the forward and the backward probabilities [Eqs.~\forwardUmomentUmain{},~\backUmeanUmain{}, and~\backUvarianceUmain{}].  
From Eqs.~\momentUSO{} and~\eqref{eq:def_pf}, we obtain Eq.~\forwardUmomentUmain{}:
\begin{align}
     \mathbb{E}_{P}[J(\tau)^r]&=\sum_{i,i^{\prime}}\int\mathcal{D}\zeta_{\tau}\,P(i,i^{\prime},\zeta_{\tau})J(\zeta_{\tau})^{r}=\sum_{i}q_i(0) \int\mathcal{D}\zeta_\tau  J(\zeta_{\tau})^{r}\braket{\zeta_\tau,i|\left(\sum_{i^\prime}\ket{i^\prime}\bra{i^\prime}\right) |\zeta_\tau,i} \nonumber\\
     &=\int\mathcal{D}\zeta_\tau J(\zeta_\tau)^r\left|  \mathfrak{L}(\tau-t_K)  \prod_{j=1}^{K} \left(L_{m_j} \mathfrak{L}(t_{j}-t_{j-1})\right)\ket{\psi(0)}\right|^2
     =\mathbb{E}_{\varnothing}[J(\tau)^r]=\mathbb{E}[J(\tau)^r].
     \label{eq:forward_moment}
\end{align}
where we use Eq.~\mainUresultUmoment{} in the last equality.
Since the number of times that channel $m$ appears in $\zeta^{\mathrm{B}}_\tau$ is equal to the number of times that $m^\prime$ appears in $\zeta_\tau$ (i.e., $N^{C}_{m} (\zeta_\tau)=N^{C}_{m^\prime} (\zeta_\tau^{\mathrm{B}})$), from Eq.~\eqref{eq:def_state_ancilla_0}, we obtain 
\begin{align}
    \mathbb{E}_{Q}[N^{C}_{m}(\tau)]&=\sum_{i,i^{\prime}}\int\mathcal{D}\zeta_{\tau}\,Q(i,i^{\prime},\zeta_{\tau}^\mathrm{B})N^{C}_m(\zeta_{\tau})=\sum_{ i^\prime} q_{i^\prime}(\tau)\int\mathcal{D}\zeta_\tau N^{C}_m(\zeta_{\tau})\braket{\zeta_\tau^{\mathrm{B}},i^\prime |\left(\sum_{i}\ket{i}\bra{i}\right)|\zeta_\tau^{\mathrm{B}},i^\prime} \nonumber\\
    &=\int\mathcal{D}\zeta_\tau N^{C}_{m^\prime} (\zeta_\tau^{\mathrm{B}})\left| \mathfrak{L}(\tau-t_K^{\mathrm{B}})  \prod_{j=1}^{K} \left(L_{m_j^{\mathrm{B}}} \mathfrak{L}(t_{j}^{\mathrm{B}}-t_{j-1}^{\mathrm{B}})\right)\ket{\psi_{\varnothing}(\tau)}\right|^2 \nonumber\\
    &=\int\mathcal{D}\zeta_\tau  N^{C}_{m^\prime} (\zeta_\tau) \left |
     \mathfrak{L}(\tau-t_K)  \prod_{j=1}^{K} \left(L_{m_j} \mathfrak{L}(t_{j}- t_{j-1})\right)\ket{\psi_{\varnothing}(\tau)}\right|^2 \nonumber\\
    &=\int\mathcal{D}\zeta_\tau  N^{C}_{m^\prime}(\zeta_\tau) \left |
     V(\tau-t_K)  \prod_{j=1}^{K} \left(L_{m_j} V(t_{j}- t_{j-1})\right)\ket{\psi(\tau)}\right|^2.
     \label{eq:mean_Q_pathint}
\end{align}
Here, the last equality follows as in Eq.~\eqref{eq:S0_relation}, and we use Eq.~\eqref{eq:psi_unitary}.
Because the last term in Eq.~\eqref{eq:mean_Q_pathint} is the result of the time evolution from $t=0$ to $\tau$ under the initial condition $\rho(\tau)$, we can identify this term with the time evolution from $t=\tau$ to $2\tau$ with the initial condition $\rho(0)$.
Therefore, we obtain
\begin{align}
     \mathbb{E}_{Q}[N^{C}_{m}(\tau)]=\mathbb{E}_{\varnothing}[N^{C}_{m^\prime}([\tau,2\tau])]=\mathbb{E}[N^{C}_{m^\prime}([\tau,2\tau])].
     \label{eq:EQ_NmC}
\end{align}
Similarly, it follows that 
\begin{align}
    \mathbb{E}_{Q}[N^{C}_{m}(\tau)N^{C}_{l}(\tau)]&=\mathbb{E}_{\varnothing}[N^{C}_{m^\prime}([\tau,2\tau])N^{C}_{l^\prime}([\tau,2\tau])]=\mathbb{E}[N^{C}_{m^\prime}([\tau,2\tau])N^{C}_{l^\prime}([\tau,2\tau])].
    \label{eq:EQ_NmC_NlC}
\end{align}
By combining these relations with $c_{m^\prime}=-c_{m}$, we obtain Eqs.~\backUmeanUmain{} and~\backUvarianceUmain{}.

We discuss the reason for performing these calculations in the system $\mathcal{S}_{\varnothing}$. For Eq.~\EPUKLUdivUmain{} to hold in the system $\mathcal{S}$, it must be defined by $V(t)^\dagger$ instead of $\mathfrak{L}(t)$ in Eq.~\defUpb{}. However, when defined in this way, Eqs.~\backUmeanUmain{} and~\backUvarianceUmain{} do not hold because $V(t)^\dagger$ is not a forward time evolution operator. 

\subsection{Proofs of Eqs.~\TURUEPUmain{} and~\SigmaUlowerboundUTUR{} \label{sec:proof_TUR}}
From the result in Ref.~\cite{Nishiyama:2020:Entropy}, the Kullback-Leibler divergence with given expectation values and variances is lower bounded by  
\begin{align}
    D(P\| Q)\geq \int_0^1 \frac{\theta \left(\mathbb{E}_{P}[J(\tau)]-\mathbb{E}_{Q}[J(\tau)]\right)^2}{(1-\theta)  \mathrm{Var}_{P}[J(\tau)] + \theta \mathrm{Var}_{Q}[J(\tau)]+\theta(1-\theta)\left(\mathbb{E}_{P}[J(\tau)]-\mathbb{E}_{Q}[J(\tau)]\right)^2 }d\theta.
    \label{eq:DPQ_lowerbound}
\end{align}
Combining the CIC for anti-symmetric counting observables [Eqs.~\forwardUmomentUmain{},~\backUmeanUmain{} and~\backUvarianceUmain{}] with 
$D(P\|Q) = \Sigma_\varnothing$ [Eq.~\EPUKLUdivUmain{}] and the definition of $\gamma(\tau)$ [Eq.~\defUgamma{}], we obtain
\begin{align}
    \Sigma_{\varnothing}(\tau)&\geq \int_0^1 \frac{\theta \mathbb{E}[J(2\tau)]^2}{ (1-\theta)\mathrm{Var}[J(\tau)] +\theta \mathrm{Var}[J([\tau,2\tau])] +\theta(1-\theta)\mathbb{E}[J(2\tau)]^2 }d\theta \nonumber\\
    &\geq  \int_0^1 \frac{\theta \mathbb{E}[J(2\tau)]^2}{  \gamma(2\tau) \mathrm{Var}[J(2\tau)] /4 +\theta(1-\theta)\mathbb{E}[J(2\tau)]^2 }d\theta=\frac{1}{g(2\tau)}\ln \frac{g(2\tau)+1}{g(2\tau)-1}=2g(2\tau)^{-1}\mathrm{arctanh}(g(2\tau)^{-1}),
    \label{eq:sigma_lb}
\end{align}
where we used
\begin{align}
    \int_{0}^{1}\frac{x}{x(1-x)+a}dx&=\frac{1}{\sqrt{4a+1}}\ln\left[\frac{\sqrt{4a+1}+1}{\sqrt{4a+1}-1}\right],
    \label{eq:integration_SM}\\
    g(\tau)&:=\sqrt{\frac{\gamma(\tau)\mathrm{Var}[J(\tau)]}{\mathbb{E}[J(\tau)]^{2}}+1}=\sqrt{R(\tau)+1}.
    \label{eq:gtau_def_SM}
\end{align}
Combining $\Sigma_{\varnothing}(2\tau)\geq \Sigma_{\varnothing}(\tau)$ with 
 $\Sigma = \Sigma_\varnothing$ [Eq.~\EPUsystemO{}] 
and replacing $2\tau$ with $\tau$, we obtain
\begin{align}
    \frac{\Sigma(\tau)}{2} \geq g(\tau)^{-1}\mathrm{arctanh} (g(\tau)^{-1}).
    \label{eq:sigma_arctanh}
\end{align}
Letting $g(\tau):=\mathrm{coth}(h(X))$ and using the definition of $h(x)$, we obtain $\Sigma(\tau) /2 \geq X$. Therefore, it follows that 
\begin{align}
    g(\tau)\geq \mathrm{coth}\left[h\left( \frac{\Sigma(\tau)}{2}\right)\right],
    \label{eq:gtau_lowerbound}
\end{align}
and this inequality completes the proof of Eq.~\TURUEPUmain{}. Applying the following relation to Eq.~\eqref{eq:sigma_arctanh}, we obtain Eq.~\SigmaUlowerboundUTUR{}.
\begin{align}
    \mathrm{arctanh}\left(\frac{1}{\sqrt{R(\tau)+1}}\right)=\frac{1}{2}\ln\left(\frac{\sqrt{R(\tau)+1}+1}{\sqrt{R(\tau)+1}-1}\right)=\ln\left(\frac{1+\sqrt{R(\tau)+1}}{\sqrt{R(\tau)}}\right)=\mathrm{arcsinh} \left(\frac{1}{\sqrt{R(\tau)}}\right).
    \label{eq:arctanh_arcsinh}
\end{align}
When $\rho$ is a stationary state $\rho_{\mathrm{ss}}$, substituting $\mathbb{E}[J(2\tau)]=2\mathbb{E}[J(\tau)]$ and $\mathrm{Var}[J([\tau,2\tau])]= \mathrm{Var}[J(\tau)]$ into the first inequality in Eq.~\eqref{eq:sigma_lb}, we obtain 
\begin{align}
    \sigma_{\varnothing}\tau&\geq \int_0^1 \frac{4\mathbb{E}[J(\tau)]^2}{\mathrm{Var}[J(\tau)]  +4\theta(1-\theta)\mathbb{E}[J(\tau)]^2 }d\theta=2\tilde{g}(\tau)^{-1}\mathrm{arctanh}(\tilde{g}(\tau)^{-1}),
    \label{eq:sigma_lb_ss}
\end{align}
where $\tilde{g}(\tau):=\sqrt{\mathrm{Var}[J(\tau)] /\mathbb{E}[J(\tau)]^2+1}$. 
Using Eq.~\eqref{eq:sigma_lb_ss}, we obtain Eq.~\TURUEPUssUmain{}. 

\section{Difference between entropy production and dynamical activity}

We will show that the thermodynamic uncertainty relation via the quantum Fisher information for entropy production is not attributable to the classical case. At the end of this section, we will provide a discussion of the difference between entropy production and dynamical activity.
Consider the auxiliary dynamics with the unchanged Hamiltonian ($H_\eta=H$) and modified jump operators as in Ref.~\cite{vu2025fundamental}:
\begin{align}
    L_{m,\eta}(t)&:=\sqrt{1+\eta l_m(t)}L_m,    \label{eq:jump_alpha_def} \\
    l_{m}(t)&:=\frac{\mathrm{Tr}[L_m \rho(t)L_m^\dagger]
    -\mathrm{Tr}[L_{m^\prime} \rho(t)L_{m^\prime}^\dagger]}{\mathrm{Tr}[L_m \rho(t)L_m^\dagger]+\mathrm{Tr}[L_{m^\prime} \rho(t)L_{m^\prime}^\dagger]}=\frac{\mathrm{Tr}[L_m \rho_{\varnothing}(t)L_m^\dagger]
    -\mathrm{Tr}[L_{m^\prime}\rho_{\varnothing}(t)L_{m^\prime}^\dagger]}{\mathrm{Tr}[L_m \rho_{\varnothing}(t)L_m^\dagger]+\mathrm{Tr}[L_{m^\prime} \rho_{\varnothing}(t)L_{m^\prime}^\dagger]}. \label{eq:l_m_def}
\end{align}
The second equality in Eq.~\eqref{eq:l_m_def} follows from Eq.~\mainUresultUmoment{} for $r=1$ and $N^{C}(\tau)=N_m(\tau)$.
Applying the quantum Cram\'er--Rao inequality~\cite{Helstrom:1976:QuantumEst} 
to the system $\mathcal{S}_{\varnothing}$, we obtain the following bound:
\begin{align}
     \frac{\mathrm{Var}_{\varnothing}[J(\tau)]}{\left(\left.\partial_\eta\mathbb{E}_{\varnothing,\eta}[J(\tau)]\right|_{\eta=0} \right)^{2}}\geq \frac{1}{I_{\varnothing}(\tau)}.
     \label{eq:quantum_Cramer_Rao}
\end{align}
Since $[H, L_{m,\eta}^\dagger(t) L_{m,\eta}(t)]=0$, applying Eq.~\ULUcommuteI{} and $[U(t),\mathfrak{L}_{\eta}(t)]=0$ repeatedly, from Eqs.~\eqref{eq:scaled_cMPS_expand} and~\eqref{eq:def_scaled_cMPS_zeta}, we obtain 
\begin{align}
   & \ket{\Psi_\eta(\tau)}=U(\tau)\int \mathcal{D}\zeta_t e^{if(\zeta_t)}\ket{\Psi_{\varnothing,_\eta}(\zeta_t, \theta)}.
    \label{eq:Phi_interaction}
\end{align}
Equation~\eqref{eq:Phi_interaction} yields
\begin{align}
   \braket{\partial_\eta \Psi_{\eta}(\tau)|\partial_\eta \Psi_{\eta}(\tau)}=\int \mathcal{D}\zeta_t \braket{\partial_\eta \Psi_{\varnothing,\eta}(\zeta_t,\theta)|\partial_\eta \Psi_{\varnothing,\eta}(\zeta_t,\theta)}=\braket{\partial_\eta \Psi_{\varnothing,\eta}(\tau)|\partial_\eta \Psi_{\varnothing,\eta}(\tau)}.
   \label{eq:PsiEta_innerproduct}
\end{align}
Similarly, we obtain $|\braket{\partial_\eta \Psi_{\eta}(\tau)| \Psi_{\eta}(\tau)}|=|\braket{\partial_\eta \Psi_{\varnothing, \eta}(\tau)| \Psi_{\varnothing,\eta}(\tau)}|$. 
From Eq.~\eqref{eq:I_varnothing_eta}, we obtain the CIC for the quantum Fisher information:
\begin{align}
    I_{\varnothing}(\tau)=I(\tau),
    \label{eq:QFI_S_S0}
\end{align}
where $I_{\varnothing}(\tau)$ and $I(\tau)$ are the quantum Fisher information at $\eta=0$.
When the original system $\mathcal{S}$ is in a stationary state $\rho_{\mathrm{ss}}$,  Ref.~\cite{vu2025fundamental} showed that the quantum Fisher information is upper bounded by
\begin{align}
    I_{\mathrm{ss}}(\tau)=\frac{\tau}{2} \sum_m\frac{\left(\mathrm{Tr}[L_m \rho_{\mathrm{ss}}L_m^\dagger]
    -\mathrm{Tr}[L_{m^\prime}  \rho_{\mathrm{ss}}L_{m^\prime}^\dagger]\right)^2}{\mathrm{Tr}[L_m  \rho_{\mathrm{ss}}L_m^\dagger]+\mathrm{Tr}[L_{m^\prime}  \rho_{\mathrm{ss}}L_{m^\prime}^\dagger]}\le \tau \frac{\sigma^2}{4\mathfrak{a}}\Phi\left(\frac{\sigma}{2\mathfrak{a}}\right)^{-2}.
    \label{eq:upperbound_QFI}
\end{align}
Here, $\Phi(x)$ denotes the inverse function of $x\tanh(x)$.
For $\eta \ll 1$, letting $\rho_{{\varnothing},\eta}(t):=\rho_{\varnothing}(t)+\eta \varphi_{\varnothing}(t)+O(\eta^2)$ with $\varphi_{\varnothing}(0)=0$, the function $\varphi_{\varnothing}(t)$ satisfies
\begin{align}
    \dot{\varphi_{\varnothing}}(t)=\sum_m\mathcal{D}\left[L_{m}\right]\varphi_{\varnothing}(t)+\sum_m l_m \mathcal{D}\left[L_{m}\right]\rho_{\varnothing}(t),
    \label{eq:varphi_evolution}
\end{align}
where we use Eq.~\LindbladUdefUSO{}.
From Eq.~\eqref{eq:jump_alpha_def} and~\eqref{eq:l_m_def}, the expectation value of a counting operator is given by 
\begin{align}
    \mathbb{E}_{\varnothing,\eta}[J(\tau)]&=\int_0^\tau \sum_m c_m \mathrm{Tr}[L_{m,\eta} \rho_{\varnothing,\eta}(t)L_{m,\eta}^\dagger]dt \nonumber\\
    &=\mathbb{E}_{\varnothing}[J(\tau)]+\eta\int_0^\tau\sum_m c_m l_m(t)\mathrm{Tr}[L_{m} \rho_{\varnothing}(t)L_{m}^\dagger]dt+\eta \int_0^\tau\sum_m c_m \mathrm{Tr}[L_{m} \varphi_{\varnothing}(t)L_{m}^\dagger]dt+O(\eta^2) \nonumber\\
    &=(1+\eta)\mathbb{E}_{\varnothing}[J(\tau)]+\eta \int_0^\tau\sum_m c_m \mathrm{Tr}[L_{m} \varphi_{\varnothing}(t)L_{m}^\dagger]dt+O(\eta^2),
    \label{eq:E_varnothing_eta_J}
\end{align}
where we use 
\begin{align}
    &\sum_m c_m l_m(t)\mathrm{Tr}[L_{m} \rho_{\varnothing}(t)L_{m}^\dagger]=\frac{1}{2}\sum_m c_m l_{m}(t) \left[\mathrm{Tr}[L_{m} \rho_{\varnothing}(t)L_{m}^\dagger]+\mathrm{Tr}[L_{m^\prime} \rho_{\varnothing}(t)L_{m^\prime}^\dagger]\right]\nonumber\\
    &=\frac{1}{2}\sum_m c_m \left[\mathrm{Tr}[L_{m} \rho_{\varnothing}(t)L_{m}^\dagger]-\mathrm{Tr}[L_{m^\prime} \rho_{\varnothing}(t)L_{m^\prime}^\dagger]\right]=\sum_m c_m \mathrm{Tr}[L_{m} \rho_{\varnothing}(t)L_{m}^\dagger]
    \label{eq:sum_cm_lm_Tr}
\end{align}
in the last  equality  in Eq.~\eqref{eq:E_varnothing_eta_J}.
Therefore, we obtain 
\begin{align}
    \left.\partial_\eta\mathbb{E}_{\varnothing,\eta}[J(\tau)]\right|_{\eta=0}=\mathbb{E}_{\varnothing}[J(\tau)]+\int_0^\tau\sum_m c_m \mathrm{Tr}[L_{m} \varphi_{\varnothing}(t) L_{m}^\dagger]dt.
    \label{eq:deriv_mean_N}
\end{align}
By using the CIC for trajectory observables [Eq.~\mainUresultUmoment{}] and quantum Fisher information [Eq.~\eqref{eq:QFI_S_S0}], and combining Eqs.~\eqref{eq:quantum_Cramer_Rao} with~\eqref{eq:deriv_mean_N}, we obtain 
\begin{align}
    I(\tau) =I_{\varnothing}(\tau)\geq \frac{\left(\mathbb{E}[J(\tau)]+\int_0^\tau\sum_m c_m \mathrm{Tr}[L_{m} \varphi_{\varnothing}(t) L_{m}^\dagger] dt \right)^{2}}{\mathrm{Var}[J(\tau)]}.
    \label{eq:TUR_EP_QFI}
\end{align}
Note that the expectation value contains an additional term.

We discuss the difference from that for the dynamical activity. In Eq.~\eqref{eq:auxiliary_rescale},  the auxiliary dynamics is equivalent to scaling time by a factor of $1+\eta$ in the system $\mathcal{S}_{\varnothing}$. The function $\rho_{{\varnothing},\eta}(t)$ can be expanded as $\rho_{{\varnothing},\eta}(t)=\rho_{{\varnothing}}((1+\eta)t)=\rho_{{\varnothing}}(t)+\eta t\dot{\rho}_{\varnothing}(t)+O(\eta^2)$. This relation yields $\varphi_{\varnothing}(t)=t\dot{\rho}_{\varnothing}(t)$ and $\left.\partial_\eta\mathbb{E}_{\varnothing,\eta}[N(\tau)]\right|_{\eta=0}=\tau \partial_\tau \mathbb{E}_{\varnothing}[N(\tau)]=\tau \partial_\tau \mathbb{E}[N(\tau)]$ from Eqs.~\mainUresultUmoment{} and~\eqref{eq:deriv_mean_N}. In contrast, Equation~\eqref{eq:varphi_evolution} is not easily solved for the entropy production. Therefore, the right-hand side of Eq.~\eqref{eq:TUR_EP_QFI} includes a coherent quantum correction term.

\section{Coherent model}

\subsection{Dynamical activity}

\begin{figure}
\centering
\includegraphics[width=0.5\linewidth]{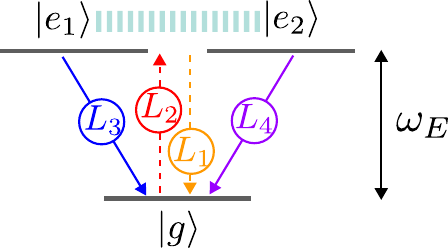}
\caption{
Schematic level diagram of the coherent model. 
The energy gap between the ground and the excited states is $\omega_E$, where the excited states $\ket{e_1}$ and $\ket{e_2}$ are degenerate. 
$L_1$ describes coherent decay from a superposition of the excited states $\ket{e_1}$ and $\ket{e_2}$ to the ground state $\ket{g}$.  
$L_2$ represents coherent excitation from the ground state $\ket{g}$ to a superposition of $\ket{e_1}$ and $\ket{e_2}$.  
$L_3$ models decay $\ket{e_1}$ to $\ket{g}$ and $L_4$ represents decay from $\ket{e_2}$ to $\ket{g}$.
}
\label{fig:coherent_model}
\end{figure}

we consider a coherent model with a degenerate energy spectrum.
Let $\ket{g}$ be the ground state and $\ket{e_1}$ and $\ket{e_2}$ be degenerate excited states (Fig.~\ref{fig:coherent_model}). The Hamiltonian is 
\begin{align}
    H=\omega_{E}(\ket{e_{1}}\bra{e_{1}}+\ket{e_{2}}\bra{e_{2}}),
    \label{eq:Deg_H_def_supp}
\end{align}
where $\omega_E > 0$ is the energy gap between the ground and the excited states. 
Here, the energy of the ground state is assumed to be $0$. 
We consider the following jump operators:
\begin{align}
L_{1} &= \sqrt{\gamma_{1}} \ket{g} \left( \bra{e_{1}} + \bra{e_{2}} \right),\label{eq:L1_coh_def} \\
L_{2} &= \sqrt{\gamma_{2}} \left( \ket{e_{1}} + \ket{e_{2}} \right) \bra{g},\label{eq:L2_coh_def} \\
L_{3} &= \sqrt{\gamma_{3}} \ket{g} \bra{e_{1}},\label{eq:L3_coh_def} \\
L_{4} &= \sqrt{\gamma_{4}} \ket{g} \bra{e_{2}},\label{eq:L4_coh_def}
\end{align}
where $\gamma_i > 0$ are transition rates. 
It is easy to check that $L_m$ in Eqs.~\eqref{eq:L1_coh_def}--\eqref{eq:L4_coh_def} satisfy Eq.~\condition{},
$[L_{1},H]=\omega_{E}L_{1}$, $[L_{2},H]=-\omega_{E}L_{2}$, $[L_{3},H]=\omega_{E}L_{3}$, and $[L_{4},H]=\omega_{E}L_{4}$.
By solving the Lindblad equation [Eq.~\LindbladUdef{}],
the steady-state density operator in the basis of $[\ket{g},\ket{e_1},\ket{e_2}]$ is given by
\begin{align}
    \rho_\mathrm{ss}=\left[\begin{array}{ccc}
\rho_{gg} & 0 & 0\\
0 & \rho_{e_{1}e_{1}} & \rho_{e_{1}e_{2}}\\
0 & \rho_{e_{1}e_{2}} & \rho_{e_{2}e_{2}}
\end{array}\right],
\label{eq:rho_ss_degenerate_supp}
\end{align}
where
\begin{align}
    \rho_{gg}&=\frac{\gamma_{1}\gamma_{3}+\gamma_{1}\gamma_{4}+\gamma_{3}\gamma_{4}}{\gamma_{1}\gamma_{3}+\gamma_{1}\gamma_{4}+\gamma_{3}\gamma_{2}+\gamma_{2}\gamma_{4}+\gamma_{3}\gamma_{4}},\label{eq:rho_gg_def}\\
    \rho_{e_{1}e_{1}}&=\frac{\left(\gamma_{3}+\gamma_{4}\right)\gamma_{2}\left(\gamma_{1}+\gamma_{4}\right)}{\left(\gamma_{1}\gamma_{3}+\gamma_{1}\gamma_{4}+\gamma_{3}\gamma_{4}\right)\left(2\gamma_{1}+\gamma_{3}+\gamma_{4}\right)}\rho_{gg},\label{eq:rho_e1e1_def}\\
    \rho_{e_{2}e_{2}}&=\frac{\gamma_{2}\left(\gamma_{3}+\gamma_{4}\right)\left(\gamma_{1}+\gamma_{3}\right)}{\left(\gamma_{1}\gamma_{3}+\gamma_{1}\gamma_{4}+\gamma_{3}\gamma_{4}\right)\left(2\gamma_{1}+\gamma_{3}+\gamma_{4}\right)}\rho_{gg},\label{eq:rho_e2e2_def}\\
    \rho_{e_{1}e_{2}}&=\frac{\left(\left(\gamma_{3}+\gamma_{4}\right)\gamma_{1}+2\gamma_{3}\gamma_{4}\right)\gamma_{2}}{\left(\left(\gamma_{3}+\gamma_{4}\right)\gamma_{1}+\gamma_{3}\gamma_{4}\right)\left(2\gamma_{1}+\gamma_{3}+\gamma_{4}\right)}\rho_{gg}.\label{eq:rho_e1e2_def}
\end{align}
We observe that the non-diagonal element $\rho_{e1e2}$ is positive when $\gamma_2 > 0$.
We define two operators that extract different components of a density matrix $\rho$: the operator $\mathcal{E}_\mathrm{d}(\rho)$ returns a matrix containing only the diagonal elements of $\rho$, while $\mathcal{E}_\mathrm{nd}(\rho)$ returns a matrix containing only the non-diagonal elements.
The dynamical activity is given by
\begin{align}
\mathfrak{a}&=\mathrm{Tr}\left[\sum_{m}L_{m}\rho_{\mathrm{ss}}L_{m}^{\dagger}\right]\nonumber\\&=\mathrm{Tr}\left[\sum_{m}L_{m}\mathcal{E}_{\mathrm{d}}(\rho_{\mathrm{ss}})L_{m}^{\dagger}\right]+\mathrm{Tr}\left[\sum_{m}L_{m}\mathcal{E}_{\mathrm{nd}}(\rho_{\mathrm{ss}})L_{m}^{\dagger}\right]\nonumber\\&=\mathfrak{a}_{\mathrm{d}}+\mathfrak{a}_{\mathrm{nd}},
    \label{eq:dynamical_activity_ss_supp}
\end{align}
where $\mathfrak{a}_\mathrm{d}$ and $\mathfrak{a}_\mathrm{nd}$ arise from the diagonal and non-diagonal elements, respectively, of $\rho_\mathrm{ss}$.
Specifically, they are given by
\begin{align}
    \mathfrak{a}_{\mathrm{d}}&=\gamma_{1}\left(\rho_{e_{1}e_{1}}+\rho_{e_{2}e_{2}}\right)+\gamma_{3}\rho_{e_{1}e_{1}}+\gamma_{4}\rho_{e_{2}e_{2}}+2\gamma_{2}\rho_{gg},\label{eq:ad_def_SM}
    \\\mathfrak{a}_{\mathrm{nd}}&=2\gamma_{1}\rho_{e_{1}e_{2}}.\label{eq:and_def_SM}
\end{align}
Since $\rho_{e_{1}e_{2}}$ is non-negative, $\mathfrak{a}_\mathrm{nd}$ is non-negative as well. 

We perform a numerical simulation for the coherent model, whose results are shown in Fig.~\FIGnumericalUsimUDA{} in the main text. Parameters are $\omega_E=1$, $\gamma_n \in (0,1)$, and $\tau \in (0.1, 10)$.

\subsection{Entropy production}

\begin{figure}
\includegraphics[width=0.5\linewidth]{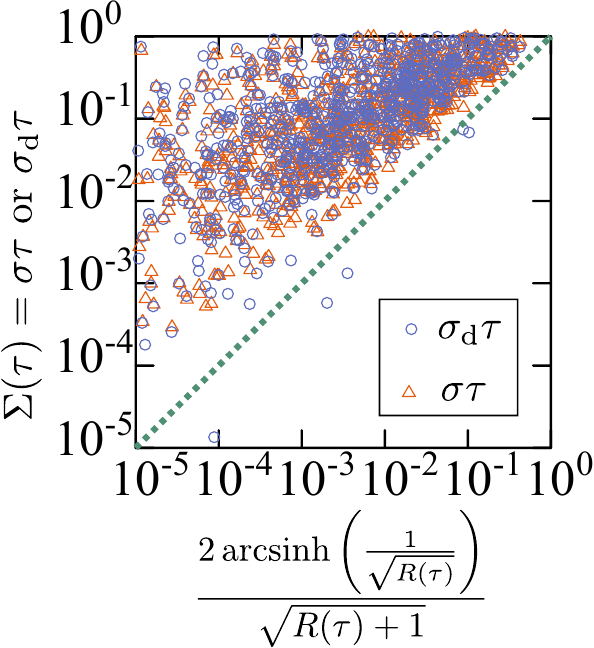}
    \caption{
The results of our numerical simulations, which validate Eq.~\SigmaUlowerboundUTUR{}. We plot the entropy production, $\Sigma(\tau)$, against the right-hand side of Eq.~\SigmaUlowerboundUTUR{} for two cases: $\Sigma(\tau) = \sigma\tau$ (red triangles) and $\Sigma(\tau) = \sigma_{\mathrm{d}}\tau$ (blue circles).
    Here, $\sigma_{\mathrm{d}}\tau$ is the diagonal contribution of the entropy production defined in Eq.~\eqref{eq:Sigma_decomposition}. The dotted line indicates the equality case of Eq.~\SigmaUlowerboundUTUR{}.
    }
    \label{fig:entropy_production_simulation}
\end{figure}

A numerical simulation is also performed for the entropy production case.
The model is essentially the same as the one used in the previous section for the dynamical activity case;
however, since every transition requires a corresponding reverse transition,
the following jump operators are used:
\begin{align}
    L_{1}&=\sqrt{\gamma_{1}}\,\ket{g}\left(\bra{e_{1}}+\bra{e_{2}}\right),\label{eq:L1_EP}\\
    L_{2}&=\sqrt{\gamma_{2}}\,\left(\ket{e_{1}}+\ket{e_{2}}\right)\bra{g},\label{eq:L2_EP}\\
    L_{3}&=\sqrt{\gamma_{3}}\,\ket{g}\bra{e_{1}},\label{eq:L3_EP}\\
    L_{4}&=\sqrt{\gamma_{4}}\,\ket{e_{1}}\bra{g},\label{eq:L4_EP}\\
    L_{5}&=\sqrt{\gamma_{5}}\,\ket{g}\bra{e_{2}},\label{eq:L5_EP}\\
    L_{6}&=\sqrt{\gamma_{6}}\,\ket{e_{2}}\bra{g}.\label{eq:L6_EP}
\end{align}
As shown in Eq.~\eqref{eq:dynamical_activity_ss_supp}, the entropy production can be divided into those arising from diagonal elements and those from non-diagonal elements:
\begin{align}
    \Sigma(\tau)&=\tau\sum_{m}\Delta s_{m}\mathrm{Tr}[L_{m}\rho_{\mathrm{ss}}L_{m}^{\dagger}]\nonumber\\&=\tau\left(\sum_{m}\Delta s_{m}\mathrm{Tr}[L_{m}\mathcal{E}_{\mathrm{d}}(\rho_{\mathrm{ss}})L_{m}^{\dagger}]+\sum_{m}\Delta s_{m}\mathrm{Tr}[L_{m}\mathcal{E}_{\mathrm{nd}}(\rho_{\mathrm{ss}})L_{m}^{\dagger}]\right)\nonumber\\&=\tau\left(\sigma_{\mathrm{d}}+\sigma_{\mathrm{nd}}\right).
\label{eq:Sigma_decomposition}
\end{align}
We perform a numerical simulation for validating Eq.~\SigmaUlowerboundUTUR{}. 
For the simulation, the values of $\gamma_n$ in Eq.~\eqref{eq:L1_EP}--\eqref{eq:L6_EP} are randomly determined, and the values of $\tau$ is also randomly chosen.
The coefficients $c_m$ necessary for the current $J(\tau)$ are randomly determined so that they satisfy the antisymmetric condition.
Note that the parameters are $\omega_E=1$, $\gamma_n \in (0,1)$, and $\tau \in (0.1, 10)$. 
Using these random values, the variance $\mathrm{Var}[J(\tau)]$ and mean $\mathbb{E}[J(\tau)]$ of the current as well as entropy production $\Sigma(\tau)$ are calculated.
We repeat this procedure many times and plot each realization in Fig.~\ref{fig:entropy_production_simulation}. 
For each random realization, $\Sigma(\tau)$,
the left-hand side of Eq.~\SigmaUlowerboundUTUR{},
is plotted against $2\mathrm{arcsinh}\left(1/\sqrt{R(\tau)}\right)/\sqrt{R(\tau)+1}$,
which is the right-hand side of Eq.~\SigmaUlowerboundUTUR{},
for two cases: $\Sigma(\tau)=\sigma\tau$ (triangles) and $\Sigma(\tau)=\sigma_\mathrm{d}\tau$ (circles). 
In Fig.~\ref{fig:entropy_production_simulation}, the dotted line denotes the equality case of Eq.~\SigmaUlowerboundUTUR{}. 
Note that the points should be above the dotted line when Eq.~\SigmaUlowerboundUTUR{} is satisfied. 
Apparently, all triangles are above the dotted line, which confirms Eq.~\SigmaUlowerboundUTUR{}.
However, when we replace the entropy production by its diagonal contribution (i.e., $\Sigma(\tau)=\sigma_{\mathrm{d}}\tau$), some circles are below the line, indicating that Eq.~\SigmaUlowerboundUTUR{} with $\Sigma(\tau)$ replaced by the diagonal contribution, is not satisfied. 
This indicates that the system's coherence improves accuracy. 

